\documentclass[11pt]{article}

\usepackage[utf8]{inputenc}
\usepackage[T1]{fontenc}
\usepackage{amsmath, amsfonts, amssymb}
\usepackage{graphicx}
\usepackage{url}
\usepackage{hyperref}
\usepackage[numbers,square]{natbib}

\usepackage{pifont}
\newcommand{\cmark}{\ding{51}}
\newcommand{\xmark}{\ding{55}}
\usepackage{listings}
\usepackage{caption}

\usepackage[nolist]{acronym} 
\acrodef{DNS}{Domain Name System}
\acrodef{DoT}{DNS over TLS}
\acrodef{DoH}{DNS over HTTPS}
\acrodef{DoQ}{DNS over QUIC}
\acrodef{ODoH}{Oblivious DNS over HTTPS}
\acrodef{AVD}{Android Virtual Devices}
\acrodef{SDK}{Software Development Kit}
\acrodef{ADB}{Android Debug Bridge}

\usepackage{setspace}
\usepackage[margin=0.7in]{geometry}
\usepackage{booktabs} 
\usepackage{multirow} 
\usepackage{multicol} 
\usepackage{xcolor}
\usepackage{colortbl}
\usepackage{longtable}
\usepackage{adjustbox}

\usepackage{algorithm}
\usepackage{algorithmic}
\usepackage{subfigure}
\usepackage{color}
\usepackage{wrapfig}

\hypersetup{
    colorlinks=true,
    linkcolor=blue,
    filecolor=magenta,      
    urlcolor=cyan,
    citecolor=blue,
}

\title{PARROT: \underline{P}ortable \underline{A}ndroid \underline{R}eproducible t\underline{r}affic \underline{O}bservation \underline{T}ool}

\author{
Andrea Jimenez-Berenguel\thanks{Corresponding author: \texttt{andrejim@pa.uc3m.es}}, Celeste Campo, Marta Moure-Garrido, Carlos Garcia-Rubio, \\Daniel Díaz-Sanchez, Florina Almenares
\\
\textit{Department of Telematic Engineering, University Carlos III of Madrid, Spain.}
}

\date{}

\begin{document}

\maketitle

\begin{abstract}
The rapid evolution of mobile security protocols and limited availability of current datasets constrains research in app traffic analysis. This paper presents PARROT, a reproducible and portable traffic capture system for systematic app traffic collection using Android Virtual Devices. The system provides automated environment setup, configurable Android versions, traffic recording management, and labeled captures extraction with human-in-the-loop app interaction. PARROT integrates mitmproxy for optional traffic decryption with automated SSL/TLS key extraction, supporting flexible capture modes with or without traffic interception. We collected a dataset of 80 apps selected from the MAppGraph dataset list, providing traffic captures with corresponding SSL keys for decryption analysis. Our comparative analysis between the MAppGraph dataset (2021) and our dataset (2025) reveals app traffic pattern evolution across 50 common apps. Key findings include migration from TLSv1.2 to TLSv1.3 protocol, with TLSv1.3 comprising 90.0\% of TCP encrypted traffic in 2025 compared to 6.7\% in 2021. QUIC protocol adoption increased substantially, with all 50 common apps generating QUIC traffic under normal network conditions compared to 30 apps in 2021. DNS communications evolved from predominantly unencrypted Do53 protocol (91.0\% in 2021) to encrypted DoT protocol (81.1\% in 2025). The open-source PARROT system enables reproducible app traffic capture for research community adoption and provides insights into app security protocol evolution.
\end{abstract}


\section{Introduction}

The rapid growth of mobile traffic data has created an increased demand for monitoring and analysis of mobile network communications. Mobile traffic analysis serves as the foundation for mobile traffic engineering, driving the need for up-to-date datasets of traffic generated by mobile applications (apps) within the research community. These datasets enable researchers to conduct network analysis, characterize app traffic patterns through fingerprinting and classification techniques, and analyze potential security vulnerabilities including private data leakage.

The mobile communication landscape has evolved substantially over the past decade, with continuous development of encryption protocols aimed at enhancing user privacy and minimizing security vulnerabilities. Traditional HTTP communications have been secured through SSL/TLS protocols integrated over TCP with HTTPS, which prevents payload inspection of encrypted packets. This security enhancement has extended to UDP communications through QUIC protocol, which incorporates TLS as an integral component while maintaining UDP's advantages of lower latency by eliminating the need for TCP connection establishment. \ac{DNS} communications have similarly evolved from unencrypted DNS over port 53 (Do53)~\cite{rfc1035dns} to encrypted variants including \ac{DoT}~\cite{rfc7858dot}, \ac{DoH}~\cite{rfc8484doh}, \ac{DoQ}~\cite{rfc9250doq}, and emerging technologies such as \ac{ODoH}~\cite{rfc9230odoh}, which provide confidentiality, authenticity, and integrity for DNS communications.

Android operating system updates have systematically enforced these security enhancements. Starting with Android 9 (API level 28), HTTP clients such as URLConnection, Cronet, and OkHttp enforce HTTPS usage by default, disabling cleartext support~\cite{android_cleartext_communications}. Android 10 (API level 29) enabled TLS 1.3 by default for all TLS connections~\cite{android_security_ssl_tls13}. QUIC protocol support was integrated through HttpEngine and Cronet network stacks~\cite{android_network_stacks}. For DNS security, Android 9 introduced \ac{DoT} transport security, followed by DNS over HTTP/3 (DoH3) support in Android 11~\cite{android_insecure_dns}.

Many researchers continue to generate their own traffic datasets due to the limited availability of public datasets that reflect current encrypted traffic patterns, provide reliable ground truth labeling, and include sufficient applications for analysis across various app categories. Among publicly available datasets with app-level labeling in PCAP format files, the most recent collection dates to 2023~\cite{Mankowski2023dataset}, and only four datasets contain more than 50 applications~\cite{Pham2021mappgraph, Zhao2022NUDTdataset, Mankowski2023dataset, Bayat2024ITC60dataset}. Additionally, many datasets cannot be released due to privacy constraints~\cite{Wang2020netlog,Rezaei2020notpublicdataset,solost} or datasets that used to be publicly available are no longer accessible~\cite{Ren2019dataset}.

The datasets collected in the literature were generated using custom traffic capture scenarios designed specifically for those collection campaigns. While some studies utilize reusable traffic capture frameworks such as MIRAGE~\cite{Aceto2019miragedataset} or NetLog~\cite{Wang2020netlog}, these systems are not publicly available as they require physical infrastructure for traffic capture. Consequently, the development of traffic capture systems has predominantly focused on creating ad hoc solutions rather than accessible reusable frameworks. These custom systems often require physical devices and substantial infrastructure, limiting their accessibility and reproducibility across different research environments. The lack of standardized capture methodologies complicates comparative analysis between studies and hinders research reproducibility.

Current datasets lack updated security protocols and do not reflect modern network communication patterns. Additionally, to the best of our knowledge, no publicly available standardized tools exist for mobile app traffic capture. These gaps in both datasets and methodologies limit researchers' ability to conduct mobile traffic analysis. Therefore, this paper presents PARROT: \underline{P}ortable \underline{A}ndroid \underline{R}eproducible t\underline{r}affic \underline{O}bservation \underline{T}ool, a portable and reproducible traffic capture system that enables mobile Android app traffic collection without requiring physical devices or extensive infrastructure. Our contributions include:

\begin{itemize}
\item Capture system framework publicly available: A complete containerized workflow for systematic app traffic capture using \ac{AVD} in a controlled environment, including automated environment setup, configurable Android versions, traffic recording management, and labeled result extraction with human-in-the-loop app interaction. The traffica capture system PARROT is available in GitHub\footnote{https://github.com/AndreaJimBerenguel/PARROT\_capture\_system\_tool}.
\item Integration of man-in-the-middle proxy in the capture system for optional traffic decryption capabilities with automated SSL/TLS key extraction, supporting flexible capture modes with or without traffic interception.
\item A publicly available dataset of 80 apps with corresponding SSL keys named PARROT2025\_mitmproxy dataset~\cite{jimenez2025parrot}. The apps where selected from the listed apps by the authors of MAppGraph dataset~\cite{Pham2021mappgraph}.
\item Comparative analysis between our dataset and MAppGraph dataset demonstrating app traffic pattern evolution between 2021 and 2025, revealing protocol adoption trends.
\end{itemize}

The remainder of this paper is organized as follows. Section~\ref{sec:related_work} reviews existing mobile traffic datasets and capture methodologies. Section~\ref{sec:capture_system} describes our proposed capture system architecture and implementation details. Section~\ref{sec:dataset_description} presents our collected dataset characteristics and traffic analysis. Section~\ref{sec:analysis_datasets} provides comparative analysis with an existing dataset to demonstrate protocol evolution trends. Section~\ref{sec:conclusion} concludes the paper and discusses future research directions.

\section{Related Work}\label{sec:related_work}

In recent years, the research community has acknowledged several public traffic datasets. In this section we review the datasets with mobile traffic apps publicly available and the capture methodology that the authors used to collect the corresponding traffic among other capture system of the literature. 

\subsection{Datasets Publicly Available}

This subsection reviews app traffic datasets that provide ground truth labeling for apps and/or app services captured in network traces, either through direct labeling or specified in accompanying logs. The majority of these datasets are available in PCAP format~\cite{Pham2021mappgraph, Zhao2022NUDTdataset, Mankowski2023dataset, Bayat2024ITC60dataset, Nikbakht2024ITC5dataset, Jiang2025dataset}, which offers researchers maximum flexibility by preserving raw traffic data that allows for custom feature extraction and analysis approaches tailored to specific research objectives. We also include three datasets that provide pre-processed traffic features in JSON format~\cite{Aceto2019miragedataset, Jiang2023dataset, Guarino2024datasetmirage}. Table~\ref{tab:datasets_summary} summarize the main characteristics of the datasets. The following datasets are presented in chronological order according to their publication dates.

\begin{table*}[ht!]
\centering
\caption{Summary of publicly available app traffic datasets.}
\label{tab:datasets_summary}
\adjustbox{width=\textwidth,center}{%
\begin{tabular}{lccccc}
\toprule
\textbf{Dataset \& Reference} & \textbf{Release year} & \textbf{Collection year} & \textbf{\# Apps} & \textbf{Format} & \textbf{Capture System} \\
\midrule
MIRAGE-2019~\cite{Aceto2019miragedataset} & 2019 & 2017-2019 & 20 & JSON & MIRAGE \\
MAppGraph~\cite{Pham2021mappgraph} & 2021 & 2021 & 101 (released version: 81) & PCAP & Custom \\
NUDT\_MobileTraffic~\cite{Zhao2022NUDTdataset} & 2022 & 2020 & 350 & PCAP & NetLog \\
Jiang et al.~\cite{Jiang2023dataset} & 2023 & 2020 & 53 & JSON & Custom \\
Mankowski et al.~\cite{Mankowski2023dataset} & 2023 & 2023 & 90 & PCAP & Custom \\
ITC-Net-Blend-60~\cite{Bayat2024ITC60dataset} & 2024 & 2021 & 60 & PCAP & Custom using PCAPdroid \\
ITC-net-audio-5~\cite{Nikbakht2024ITC5dataset} & 2024 & not specified & 5 & PCAP & Custom using PCAPdroid \\
MIRAGE-APPxACT-2024~\cite{Guarino2024datasetmirage} & 2024 & 2021-2023 & 20 & JSON & MIRAGE \\
Jiang et al. BRAS~\cite{Jiang2025dataset} & 2025 & not specified & 9 (services) & PCAP & Custom \\
Jiang et al. ONU~\cite{Jiang2025dataset} & 2025 & not specified & 9 (services) & PCAP & Custom \\
\bottomrule
\end{tabular}
}
\smallskip

\textbf{\# Apps:} Number of apps. \textbf{Format:} PCAP = Raw packet capture files, JSON = Pre-processed features. \textbf{Capture System:} Custom = Custom capture system, NetLog/MIRAGE/PCAPdroid = Literature systems.
\end{table*}

The MIRAGE-2019 dataset~\cite{Aceto2019miragedataset} contains traffic from 20 apps with features extracted at the biflow level stored in JSON files~\footnote{https://traffic.comics.unina.it/mirage/mirage-2019.html}. Each biflow is labeled with the corresponding Android package-name. The capture period was from May 2017 to May 2019 in an Italian university. They employed different mobile devices operated by volunteers to generate the traffic. Traffic was generated through the MIRAGE capture system. 

The MAppGraph dataset~\cite{Pham2021mappgraph} from 2021 provides traffic captures from 101 Android apps popular in Vietnam and is available upon request; however, the version shared with researchers includes 81 apps~\footnote{https://soeai.github.io/mappgraph/}. It offers a broad range of app traffic and an average of 330 minutes of traffic per app. The app traffic was collected with a custom capture system and stored in PCAP files. The resulting dataset contains traffic generated by human volunteers using smartphones and running the predefined app in a controlled university environment. 

The NUDT\_MobileTraffic dataset~\cite{Zhao2022NUDTdataset} was collected in 2020 and contains traffic from 350 Chinese apps. The dataset is anonymized and requires direct contact with authors for access~\footnote{https://github.com/Abby-ZS/NUDT\_MobileTraffic}. The authors used the capture system Netlog to collect the traffic in PCAP files and label it at the app level. The traffic was generated with smartphones operated by volunteers installing and running the designed app. The PCAP files are not labeled by app but each record in the log file contains a timestamp, an app label (in chinese), and the five-tuple information (Protocol, SrcIP, SrcPort, DstIP, DstPort) of one Biflow.

The dataset released in 2023 by Jiang et al.~\cite{Jiang2023dataset} contains traffic from 53 apps selected from the list of apps used by AppScanner~\cite{Taylor2016appscanner}. The traffic was collected in 2020 across different mobile platforms by volunteers using a custom capture system. The dataset includes 5 subsets with 50 JSON files labeled per app~\footnote{https://github.com/jmhIcoding/fdan}. Three of these subsets have regional gateways in Hong Kong and the other two subsets have regional gateways in California. 

Mankowski et al.~\cite{Mankowski2023dataset} collected a dataset in 2023 containing traffic from 90 most popular Android apps in Germany~\footnote{https://zenodo.org/records/7950522}. They focus on game category and general-purpose app category. It contains a PCAP file per app with approximately five minutes of duration. The dataset was generated with a custom capture system and it provides traffic captures generated through human interaction.

The ITC-Net-Blend-60 dataset~\cite{Bayat2024ITC60dataset} collected in 2021 contains 1,159 PCAP traces from 60 Android apps from Google Play Store and two major Iranian Android app markets totaling 36 GB of traffic data~\footnote{https://data.mendeley.com/datasets/ssv23kfcgs/3}. The authors integrated PCAPdroid in the capture system to filter the traffic generated by the app from the background traffic. Traffic was generated by five volunteers interacting with apps for 3 to 15 minutes across five different network scenarios.

The ITC-net-audio-5 dataset~\cite{Nikbakht2024ITC5dataset} focuses specifically on audio streaming apps, containing 500 PCAP files from 5 trending apps (Google Meet, Skype, Telegram, WhatsApp, and SoundCloud) with each file lasting 3 to 4 minutes. The authors do not specified the collection year of the dataset but it was release in 2024. The dataset was collected in a university of Iran integrating PCAPdroid to filter the app-specific traffic, as the previous dataset ITC-Net-Blend-60~\cite{Bayat2024ITC60dataset}.

The MIRAGE-APPxACT-2024 dataset~\cite{Guarino2024datasetmirage} was captured with an updated version of the MIRAGE capture system from April 2021 to December 2023. The traffic was generated by volunteers using mobile devices as the MIRAGE-2019 dataset~\cite{Aceto2019miragedataset}. The dataset is composed by traffic from 20 popular apps. The resulting data is stored in JSON files labeled by app containing information at the biflow level~\footnote{https://traffic.comics.unina.it/mirage/mirage-2024.html}. 

In 2025 two dataset were released by Jiang et al.~\cite{Jiang2025dataset}. The capture system consisted on traffic acquisition from access devices, specifically the Broadband Remote Access Server (BRAS), and edge devices, particularly the Optical Network Unit (ONU). They captured traffic from real users in China but it is not specified the collection date period. Upon completion of the capture process, the network traffic were classified and labeled in nine standardized service app categories. The authors released two anonymized datasets, the one captured from BRAS network units with a total of 23 PCAP files and a size of 5.36GB, and the one captured from ONU network units with a total of 41 data files and a size of 5.05GB~\footnote{https://springernature.figshare.com/articles/dataset/Tracffic\_data\_from\_real\_network\_environment/28380347}. Both datasets cover 7 business app categories and each PCAP file is labeled with the corresponding category. 

Among the datasets reviewed, the most recent traffic captures were collected in 2023, highlighting a gap in up-to-date app traffic data. Furthermore, only four datasets provide traffic traces from more than 50 apps in PCAP format: MAppGraph (101 apps), NUDT\_MobileTraffic (350 apps), Mankowski et al. (90 apps), and Jiang et al. 2023 (53 apps). The PCAP format is particularly valuable as it preserves raw traffic data, enabling researchers to develop models based on any traffic object (e.g., packet, flow, and bag of flows), feature extraction method, or innovative analytical approaches. This indicates a need for updated app traffic collections that can support research methodologies and reflect current app behaviors and security implementations.

\subsection{Capture Systems Overview}

The methodologies employed for app traffic capture vary significantly across research efforts. Most of them are custom capture systems and they do not provide all the details of the workflow for the traffic capture. First, we present the custom capture systems that use physical devices to capture traffic, then we present a capture system methodology that use non-physical devices, they use an emulator to generate the traffic. Then, the capture systems previously available in the literature and finally PCAPdroid which is a capture tool publicly available. Table~\ref{tab:capture_systems_summary} summarized the summarize the main characteristics of the capture systems.

\begin{table*}[ht!]
\centering
\caption{Summary of app traffic capture systems.}
\label{tab:capture_systems_summary}
\adjustbox{width=\textwidth,center}{%
\begin{tabular}{lccccccc}
\toprule
\textbf{Name \& Ref} & \textbf{Release Year} & \textbf{Scenario} & \textbf{Traffic Nature} & \textbf{Labels} & \textbf{TLS keys} & \textbf{Availability} \\
\midrule
\rowcolor{pink!15} MIRAGE~\cite{Aceto2019miragedataset,Guarino2024datasetmirage} & 2019 & Physical & Human-generated & Post-processing & \xmark & Only Description \\
\rowcolor{pink!15} NetLog~\cite{Wang2020netlog} & 2020 & Physical & Human-generated & App isolation & \xmark & Only Description \\
\rowcolor{gray!20} Custom~\cite{Pham2021mappgraph} & 2021 & Physical & Human-generated & App isolation & \xmark & Only Description \\
\rowcolor{yellow!20} PCAPdroid tool~\cite{faranda2025pcapdroid} & 2021 & Supports both & Human-generated & App isolation & \cmark & App \\
\rowcolor{gray!15} Custom~\cite{Jiang2023dataset} & 2023 & Physical & Human-generated & Post-processing & \xmark & Only Description \\
\rowcolor{gray!15} Custom~\cite{Mankowski2023dataset} & 2023 & Physical & Human-generated & App isolation & \xmark & Only Description \\
\rowcolor{gray!15} Custom using PCAPdroid~\cite{Bayat2024ITC60dataset,Nikbakht2024ITC5dataset} & 2024 & Physical & Human-generated & Post-processing & \xmark & Only Description \\
\rowcolor{gray!15} Custom~\cite{Jiang2025dataset} & 2025 & Physical & Human-generated & Post-processing & \xmark & Only Description \\
\rowcolor{green!12} Custom~\cite{solost} & 2025 & Emulator & Script-generated & App isolation & \xmark & Code of alternative infrastructure  \\
\bottomrule
\end{tabular}
}
\smallskip

\textbf{Color coding:} \colorbox{pink!15}{Literature systems} \colorbox{gray!15}{Custom using physical devices} \colorbox{yellow!20}{Publicly available tool} \colorbox{green!12}{Custom using emulated devices}

\textbf{Scenario:} Physical = Real devices, Emulator = Virtual devices. \textbf{Traffic Nature:} Human-generated/Script-generated traffic. \textbf{Labels:} App isolation = Traffic labeled through device/app isolation, Post-processing = Labels assigned via trace analysis. \textbf{TLS Keys:} TLS session keys provided = \cmark/\xmark. \textbf{Availability:} Only Description = Methodology described only, Code = Implementation available, App = app available.
\end{table*}

\textbf{Capture systems with physical devices:} The MAppGraph dataset~\cite{Pham2021mappgraph} employed a controlled university environment using physical Android smartphones. The capture system utilized 8 smartphones connected to a WiFi router with traffic mirroring capabilities to a desktop storage system. It involved a team of 10 volunteer students conducting data collection sessions over 6 months, with each session lasting 3-4 hours. During each session, volunteers operated smartphones with only one target app installed (apart from core Android OS apps), ensuring clean app-specific traffic labeling based on device IP addresses. The system captured traffic from multiple devices simultaneously. The human-generated traffic approach aimed to reflect realistic user behavior patterns.

Mankowski et al.~\cite{Mankowski2023dataset} employed a multi-step automated pipeline for collecting traffic from the Android apps. The system used a Google Play Scraper implemented with Selenium to retrieve app IDs by scrolling through Google Play's Top charts subcategories. App downloads were performed using Raccoon APK Downloader, which retrieves apps directly from Google Play to a laptop. Traffic capture utilized a Android phone connected to a laptop via \ac{ADB} and hotspot for traffic interception. Each app was installed individually and the authors interacted with the app for 5 minutes. Then the app was uninstalled before proceeding to the next app. Traffic recording began immediately after app launch and stopped before uninstallation, creating separate PCAP files for each app.

The ITC-Net-Blend-60 dataset~\cite{Bayat2024ITC60dataset} and ITC-net-audio-5 dataset~\cite{Nikbakht2024ITC5dataset} employed similar dual-capture methodologies using Wireshark and PCAPdroid to isolate app-specific traffic from background noise. Both systems utilized a laptop or Windows PC running Wireshark connected to a smartphone or Android tablet with PCAPdroid via internet sharing through the PC's network adapter. Wireshark captured complete network data while PCAPdroid simultaneously captured app-specific traffic on the source device. Traffic filtering was performed using Python code to identify and retain only packets identical between both captures, eliminating background traffic and packets modified by PCAPdroid at network layers 3 and 4. The traffic from ITC-Net-Blend-60 dataset was generated by volunteers who interacted with apps, while the ITC-net-audio-5 dataset did not specify the interaction methodology for traffic generation.

The capture system that generated the dataset of Jiang et al.~\cite{Jiang2023dataset} utilized three mobile devices with ten participants. Participants created virtual accounts to avoid privacy issues and performed regular user interactions combined with automation operations on the latest app versions. Network traffic between smartphones and WiFi access points was passively collected and transmitted to the researchers. 

The Jiang et al.~\cite{Jiang2025dataset} capture system employed an in-situ network deployment approach collecting real Internet traffic through sensors on network access equipment. Traffic was collected from two primary components: traffic acquisition from Broadband Remote Access Server (BRAS) access devices and Optical Network Unit (ONU) edge devices within actual network infrastructure. For edge devices, the system deployed ONU plugins using port replication to collect raw traffic. And, for access devices, software packages were deployed on BRAS equipment handling 19,417 online users generating 45 Gbit/s bidirectional traffic during peak hours. The system employed deep packet inspection technology using DNS domain name resolution methods and payload byte sequence analysis to identify app protocols. They stored the collected traffic on cloud servers after data cleaning, filtering, labeling, and format conversion to generate PCAP and CSV files.

\textbf{Capture system using emulated devices:} So et al.~\cite{solost} proposed two infrastructures: their original capture system based on Android Studio Emulators running Android 11 (API level 30) in rooted mode, and an alternative using Android Cuttlefish virtual devices. The later is publicly available\footnote{https://zenodo.org/records/15128091}. Both infrastructures employed a hybrid containerized approach where each emulator connected to dedicated containerized HTTPS man-in-the-middle proxies based on mitmproxy~\cite{mitmproxy} and containerized DNS servers configured as bind9 forwarding resolvers. Traffic capture sessions were fully automated using a custom version of DroidBot for UI-guided test input generation, with each app explored for exactly three minutes per session. The system employed Frida instrumentation to disable certificate pinning. Traffic attribution relied on app isolation methodology. They also used special non-existent domain queries served as traffic delimiters to mark precise time windows for accurate traffic attribution. Network traffic was stored in PCAP format labeled by the package name.

\textbf{Capture systems available in the literature:} The NetLog capture system~\cite{Wang2020netlog} was used to collect the NUDT\_MobileTraffic dataset~\cite{Zhao2022NUDTdataset}. NetLog is an Android monitoring tool built upon the VPNService API provided in the Android \ac{SDK}. After the installation and launch on an Android device, NetLog creates a virtual network interface that enables traffic duplication at userspace without requiring root permissions. The system leverages Android's UID mapping capabilities combined with the PackageManager API to accurately associate network flows with specific apps for automated labeling. NetLog captures traffic before encapsulation and forwarding by the VPNService, ensuring no additional bias is introduced during the collection process. The capture campaign involved distributing NetLog to volunteers who were asked to install and operate the designated apps on their smartphones. The system automatically uploads raw packets and app-flow mapping files to a cloud server via WiFi.

The MIRAGE capture system~\cite{Aceto2019miragedataset,Guarino2024datasetmirage} employs a workstation-based capture server equipped with an IEEE 802.11 access point providing connectivity to rooted Android devices operated by human experimenters. The system captures traffic via \texttt{tcpdump} on the server's wired interface while simultaneously logging network system calls using strace through \ac{ADB} connections via USB hub. Traffic filtering based on device MAC addresses enables simultaneous multi-device capture without ambiguity. The system operates in three phases: capture phase collecting PCAP traces and strace log files, ground truth building phase segmenting traffic into biflows and labeling them using Android package names extracted from system calls, and dataset extraction phase producing statistical features for analysis. The enhanced MIRAGE architecture for the MIRAGE-APPxACT-2024 dataset~\cite{Guarino2024datasetmirage} upgraded the access point to IEEE 802.11n standard for improved performance and replaced strace with netstat commands for more reliable socket-to-package mapping. The enhanced system achieves exact labeling for approximately 78\% of biflows and incorporates manual activity labeling based on user interactions during capture sessions.

\textbf{Traffic capture tool publicly available:} PCAPdroid~\cite{faranda2025pcapdroid}. It is an open-source network capture and monitoring tool that operates without root privileges by leveraging Android's VpnService API to simulate a VPN for traffic interception. The tool captures network traffic from both user and system apps, enabling traffic filtering, HTTPS/TLS decryption for specific apps, and PCAP file generation. PCAPdroid processes traffic locally without creating external VPN connections and supports real-time streaming to remote receivers. However, non-root operation introduces traffic modifications at L3 and L4 layers, including synthetic IP and TCP/UDP headers for incoming packets, altered packet sizes due to L4 socket proxying, and potential disabling of certain IP and TCP features. Root mode operation provides unmodified 1:1 traffic visibility but requires elevated privileges. The tool supports TLS decryption capabilities for protocol analysis and security assessment, though limitations exist for QUIC traffic decryption and STARTTLS protocols. The first release of the tool in Google Play was on 2019 but the latest update at the time of publication of this paper, was on 2025.

This review of app traffic datasets and capture systems reveals limitations in current research resources. Among available datasets, only four provide traffic traces from more than 50 apps in PCAP format, with the most recent captures dating to 2023. This scarcity of datasets in flexible formats constrains researchers' ability to develop and evaluate traffic analysis techniques.

The capture systems proposed in the literature and used to generate these datasets are predominantly ad hoc solutions developed for specific research objectives rather than reusable frameworks. While systems like MIRAGE and NetLog have been proposed, they are not publicly available and require infrastructure for traffic collection. PCAPdroid represents the most accessible option, gaining popularity in recent years as a publicly available tool. So et al.~\cite{solost} also released an infrastructure using Cuttlefish emulator that enables traffic capture with reduced infrastructure requirements compared to their original system.

To address these limitations, we present PARROT, a complete traffic capture system that provides several advantages over existing approaches. Unlike PCAPdroid, our system preserves original packet headers and provides a complete capture workflow including isolated environment setup, automated app installation, traffic capture management, and systematic data labeling. In contrast to So et al.~\cite{solost}, we release a fully integrated system with Android Studio Emulator support, enabling researchers to configure desired Android versions within a completely containerized, portable, and reproducible environment as the authors of~\cite{solost} only containerized the mitmproxy and a DNS server. Our approach combines human-generated traffic patterns with automated technical processes, providing researchers with a complete solution for creating labeled app traffic datasets.

We also present a dataset collected with our capture system using mitmproxy mode, providing SSL key files to decrypt traffic communications. We release a dataset with traffic from 80 apps selected from the MAppGraph dataset list, delivered in PCAP format to provide flexibility for future research.

\section{Description of the Capture System}\label{sec:capture_system}

In this section, we describe our novel traffic capture system designed to address the collection of up-to-date app traffic datasets. 
We offer a complete, reproducible environment that researchers can use to create their own app-labeled datasets with decryption capabilities. Our system employs a semi-automated approach that balances automation and manual interaction with the device. We automate technical aspects such as capture labeling and storage management while we preserve human interaction with apps. This design choice deliberately avoids fully automated scripts that interact with the device as we want to reflect genuine user behavior.

\subsection{System Architecture}

\begin{figure}[htbp]
    \centering
    \includegraphics[width=0.8\columnwidth]{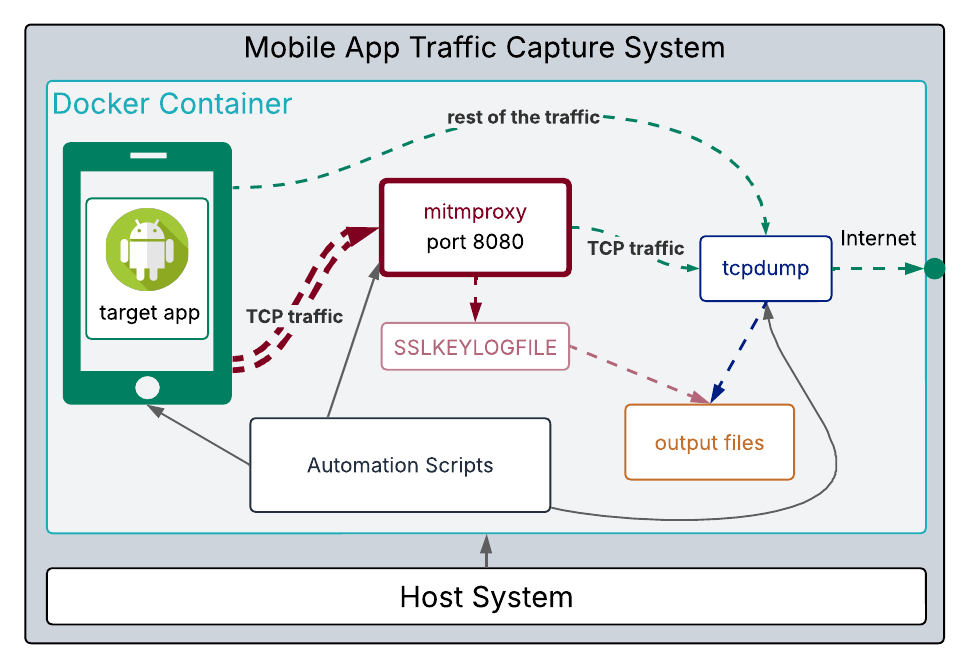}
    \caption{Architecture of the mobile traffic capture system showing the main components within the Docker container and the data flow during traffic capture.}
    \label{fig:system-architecture}
\end{figure}

Our capture system employs a containerized approach to ensure reproducibility and portability across research environments. Fig.~\ref{fig:system-architecture} illustrates the main components and data flow of our system. The system consists of a single Docker container that hosts an \ac{AVD} and all necessary traffic interception tools. By utilizing a container architecture, we are able to capture clean network traffic that is isolated from the rest of the host computer.

The principal components of our capture system are the device that generates the traffic, the tool that intercepts encrypted traffic, the tool for capturing and storing the traffic, and finally, scripts that automate the workflow.

\textbf{\textbf{Device:}} We utilize an \ac{AVD}, specifically an Android Emulator with Android 15 (API level 35), to create a controlled environment that closely mimics a real mobile device. The emulator is configured with Google APIs to provide a realistic environment for apps that require Google services.

\textit{\textbf{Traffic Intercepting:}} We employ mitmproxy~\cite{mitmproxy} as a man-in-the-middle proxy to intercept encrypted traffic over TCP. With this tool, we can intercept and extract the decryption keys from encrypted communications transmitted over TCP, not limited to HTTPS traffic. We are also able to decrypt \ac{DoT} traffic.

\textit{\textbf{Traffic Capturing:}} For packet capture, we utilize \texttt{tcpdump} to record all network traffic generated by the apps running in the AVD, ensuring that both proxy and non-proxy traffic is captured. It should be noted that the captures exclude packets transmitted between the device and the mitmproxy itself.

\textit{\textbf{Automation Scripts:}} A suite of bash scripts manages the full capture workflow, from environment setup to traffic recording and result extraction. This facilitates and expedites large-scale traffic capturing, thereby minimizing human interaction to only essential elements.

\subsection{System Workflow}

The traffic capture process follows a systematic workflow that ensures consistent and reliable data collection. First the environment is initialize 

\begin{enumerate}
    \item \textbf{Environment Initialization:} The Docker container is built with all necessary components using a Dockerfile and docker-compose configuration. This ensures identical environments conditions across capture session collection.
    
    \item \textbf{AVD Configuration:} An \ac{AVD} is launched with a pre-configured proxy setup (pointing to the internal mitmproxy) and a writable system partition that allows for certificate installation.
    
    \item \textbf{Certificate Installation:} The mitmproxy certificate is automatically installed in the system's trusted certificate store, allowing for transparent interception of TLS/SSL traffic.
    
    \item \textbf{App Installation:} The target app is installed from a local APK file. 
    
    \item \textbf{Traffic Capture Initiation:} When the capture process begins, \texttt{tcpdump} starts recording all network packets, while the \texttt{SSLKEYLOGFILE} environment variable ensures TLS session keys are recorded.
    
    \item \textbf{Human Interaction:} A human operator interacts with the app, ensuring realistic behavior patterns. This distinguishes our approach from automated testing frameworks that may not generate representative traffic.
    
    \item \textbf{Capture Termination:} Upon completion, the capture process is stopped and the capturing setup is reset. 
    
    \item \textbf{Data Extraction and Labeling:} The system automatically extracts the PCAP files and SSL key logs, labeling them with the app name, timestamp, and duration to maintain a clean, well-organized dataset.
\end{enumerate}

After describing the core components and workflow of our capture system, the next subsections present the technical details of each component and how they are implemented. This approach enables reproducible traffic capture while maintaining flexibility for different research scenarios.

\subsection{Docker Container Configuration}

Our traffic capture system is built using Docker technology, leveraging both a Dockerfile and Docker Compose to create a reproducible and portable environment. In this subsection, we review the technical details of the container and environment setup. 

The Dockerfile establishes an Ubuntu-based container that serves as the foundation for our capture environment. During the build process, the system provisions the container with essential components including the mitmproxy binary and Android \ac{SDK} command-line tools~\cite{androidsdk}. Critical dependencies are installed, encompassing \texttt{tcpdump} for packet capture, OpenJDK 17 for Android compatibility, networking utilities for system management, and X11 support to enable GUI display forwarding from the container to the host system.

The Dockerfile also automates the download and installation of Android \ac{SDK} packages necessary for emulator functionality, including the emulator binary, platform tools for device communication, Android platform specifications, and system images with Google APIs support for the target architecture. The \ac{AVD} is pre-configured during the build process as describe in~\ref{Android_emulator_setup}. Network ports 8080 and 8081 are exposed to allow researchers to monitor intercepted traffic through \texttt{localhost:8081}. 

Upon container initialization, the \texttt{start.sh} script automatically executes to launch both mitmproxy and the Android emulator with the necessary configurations.

The Docker Compose manages the orchestration and runtime configuration of our containerized environment. It provides dynamic build arguments for mitmproxy and Android SDK paths, providing flexibility for different deployment scenarios. We employ host networking to enable direct network interface access for capturing traffic generated by the AVD from within the container. While bridge mode networking would provide better container isolation, it would only capture loopback calls when attempting traffic capture inside the container, thus reducing the system's effectiveness for our functional requirements. 

Performance optimization is achieved through KVM hardware acceleration by mounting \texttt{/dev/kvm}, while GUI functionality is enabled by mounting the X11 socket, allowing the Android emulator interface to display on the host system for interactive user sessions. The container runs in privileged mode, which is necessary for Android emulator hardware acceleration.

Finally, related to traffic capture configuration, we configured the following parameters in the Docker Compose file to leverage traffic capture capabilities. In order to  decrypting encrypted traffic during traffic analysis, the \texttt{SSLKEYLOGFILE} environment variable is configured to enable automatic SSL/TLS key logging. Volume mappings establish persistent data flow between the container and host system, ensuring that captured traffic files are stored in the designated output directory on the host machine. And lastly, for the DNS Resolution we use Google Public DNS servers (8.8.8.8 and 8.8.4.4) to ensure consistent and reliable domain name resolution throughout traffic capture sessions, preventing network-related inconsistencies that could affect data consistency or quality.

\subsection{Android Emulator Setup} \label{Android_emulator_setup} 

During the container building process, we download and install the Android command-line tools, which allow us to use the Android \ac{SDK} that includes an Android device emulator. The Android Emulator simulates Android devices without requiring a physical device. We configure the Android Emulator with parameters to support traffic capture operations while maintaining mobile device behavior. We deploy a capture system environment as automatically as possible while allowing human interactions to capture realistic behavior. In this subsection, we describe the configuration of the different components used to create our emulator device.

The SDK Manager is a command-line tool that enables package management for the Android SDK. We use the \texttt{sdkmanager} to install components required for our emulator environment. Our system is configured with Android 15 (API level 35) to demonstrate the latest Android capabilities, though researchers can modify the configuration to target their preferred Android version by adjusting the platform and system image packages. The installation process includes the following packages: \texttt{emulator} provides the QEMU-based device-emulation tool for debugging and testing apps in an Android runtime environment; \texttt{platform-tools} includes essential tools such as \ac{ADB} for managing emulator instances and installing APKs; \texttt{platforms;android-35} provides the Android 15 platform support required for the target API level; and \texttt{system-images;android-35;google\_apis;x86\_64} supplies the system image with Google APIs integration for \texttt{x86\_64} architecture, enabling apps that depend on Google services to function properly within the emulated environment.

The AVD Manager is a tool that helps create and manage \ac{AVD} from the command line. We use the \texttt{avdmanager} to create our virtual device with the command:

\begin{lstlisting}
RUN ./avdmanager create avd
    -n avd-android-35-withoutplaystore 
    -k "system-images;android-35;google_apis;x86_64" 
    --device "pixel_7" 
    --force
\end{lstlisting}

The \texttt{-n} parameter specifies the AVD name as \texttt{avd-android-35-withoutplaystore}, indicating the Android version and the absence of Google Play Store because we automated the installation of the APKs via command line using the adb tool. The \texttt{-k} parameter defines the system image key that corresponds to our installed system image with Google APIs support. The \texttt{--device} parameter selects the Pixel 7 hardware profile, which provides predefined hardware characteristics including screen size, resolution, RAM, and sensor configurations that simulate a real Pixel 7 device. The \texttt{--force} flag overwrites any existing AVD with the same name, ensuring a fresh configuration for each deployment.

The emulator is launched using specific parameters to optimize traffic capture capabilities. The following launch command configures the emulator for our capture system requirements.

\begin{lstlisting}
./emulator 
    -avd avd-android-35-withoutplaystore 
    -http-proxy http://127.0.0.1:8080 
    -no-snapshot-load 
    -dns-server 8.8.8.8 
    -netdelay none 
    -netspeed full 
    -verbose 
    -show-kernel
\end{lstlisting}

The \texttt{-avd} parameter specifies which virtual device to launch. The \texttt{-http-proxy} parameter routes all TCP traffic through our mitmproxy instance running on localhost port 8080, enabling traffic interception. The \texttt{-no-snapshot-load} parameter performs a cold boot, ensuring a clean state for each capture session by preventing the emulator from loading previous session states. The \texttt{-dns-server} parameter configures Google's public DNS server (8.8.8.8) for consistent domain name resolution. The \texttt{-netdelay none} and \texttt{-netspeed full} parameters eliminate artificial network throttling to capture traffic at maximum available speeds. The \texttt{-verbose} and \texttt{-show-kernel} parameters enable detailed logging for debugging and monitoring emulator behavior during traffic capture operations.

The \ac{ADB} serves as the primary interface for managing and configuring the Android emulator throughout the traffic capture process. As a command-line tool that provides access to Android device functionality, adb enables our system to perform system-level modifications, app management, and emulator control operations from automated scripts, making our traffic capture system as automated as possible. Our system utilizes adb for four main categories of operations: system status monitoring and control, obtaining root privileges through \texttt{./adb root}, and managing emulator restarts via \texttt{./adb reboot}; file system operations such as transferring files between host and emulator using \texttt{./adb push} and remounting the system partition as writable with \texttt{./adb remount}; security configuration modifications including disabling Android's dm-verity protection system with \texttt{./adb disable-verity} to allow system partition modifications and facilitate certificate installation; and app lifecycle management through APK installation, uninstallation, and cache clearing operations to ensure clean testing environments.

\subsection{Mitmproxy Configuration}

Our capture system provides optional man-in-the-middle proxy interception through mitmproxy~\cite{mitmproxy}, which can decrypt TCP encrypted traffic when enabled via a configuration flag during scenario deployment. This configuration implements mitmproxy in regular proxy mode, the most reliable setup for intercepting app traffic when the client can be configured to use an HTTP proxy.

The mitmproxy service is initialized using the \texttt{mitmweb} variant, which provides both proxy functionality and a browser-based graphical interface for traffic monitoring. The service is launched with the command: 
\begin{lstlisting}
mitmweb 
    --listen-host 0.0.0.0 
    --listen-port 8080 
    --web-host 0.0.0.0 
    --web-port 8081
\end{lstlisting}

The \texttt{--listen-host 0.0.0.0} parameter configures the proxy to accept connections from any network interface within the container, and \texttt{--listen-port 8080} sets the proxy port that the Android emulator will use. The \texttt{--web-host 0.0.0.0} and \texttt{--web-port 8081} parameters configure the web interface to be accessible from the host system, allowing researchers to monitor intercepted traffic in real-time through a browser interface at \texttt{localhost:8081}. The process runs in the background with output redirected to \texttt{/dev/null}, and the process ID is captured for later process management during capture termination.

Mitmproxy's traffic interception capability relies on establishing itself as a trusted Certificate Authority within the Android system. The certificate download process begins by accessing mitmproxy's built-in certificate distribution service using: 

\begin{lstlisting}
curl 
    --proxy http://127.0.0.1:8080 http://mitm.it/cert/cer > /opt/android-sdk/ca.cer
\end{lstlisting}

The \texttt{mitm.it} domain is a endpoint provided by mitmproxy that serves the current session's CA certificate when accessed through the proxy itself. This approach ensures that the correct certificate is obtained for the running mitmproxy instance. If the certificate download fails, the system terminates with an error message to prevent incomplete configuration that would result in certificate validation failures during traffic capture.

The certificate installation process leverages the system preparations made through adb configuration to establish mitmproxy as a trusted Certificate Authority. The configuration process involves several steps:

\begin{lstlisting}
./adb push /opt/android-sdk/ca.cer /sdcard/Download/ca.cer
./adb root
./adb disable-verity
./adb reboot
\end{lstlisting}

First, the downloaded certificate file is transferred to the emulator's accessible storage location; second, administrative access is obtained to enable system-level modifications necessary for certificate trust store manipulation; third, Android's dm-verity verification system is disabled, which normally protects the system partition from unauthorized modifications. This step is required to allow certificate installation in the system trust store rather than the user trust store. Fourth, the emulator is rebooted to apply the dm-verity changes and reinitialize the system with the modified security settings. Following reboot completion, the certificate installation is completed manually through the Android Settings interface as describe in~\ref{traffic_capture_process}. 

\subsection{Traffic Capture Process}\label{traffic_capture_process}

The traffic capture process represents the operational phase of our system, where actual app traffic is recorded and analyzed. This process combines automated setup procedures with human interaction to generate realistic traffic patterns while maintaining systematic data collection and labeling.

The system supports automated APK installation through the \texttt{installation\_and\_traffic\_capture.sh} script, which uses adb commands to install apps on the emulator. The script accepts three parameters: a mandatory app name parameter that specifies the target app for traffic capture, an optional \texttt{-p} flag followed by the APK file path for installation, and an optional \texttt{-u} flag to uninstall the app after capture completion. APK files can be obtained from various sources including third-party repositories such as APKMirror, APKPure, and Up2Down\footnote{https://www.apkmirror.com/, https://apkpure.com/es/, https://en.uptodown.com/}, which, while not official Google Play Store sources, we can automate the installation process with this apps packages. When the \texttt{-p} parameter is provided, the system executes the following command to install the target app.
\begin{lstlisting}
./adb -s "\$EMULATOR\_ID" install "/tmp/\$\{NAME\_APK\}.apk"
\end{lstlisting}
The optional nature of the installation and uninstallation parameters provides flexibility for different capture scenarios. When users want to capture traffic from the same app across multiple sessions, they can omit the \texttt{-p} parameter if the app is already installed from a previous capture process, eliminating unnecessary reinstallation steps. Similarly, the optional \texttt{-u} flag allows users to maintain the app on the emulator between capture sessions, enabling continuous testing scenarios or comparative analysis across different usage patterns without the overhead of repeated installation and configuration processes. Before installation, the system performs cache clearing operations for common apps like Chrome to ensure clean testing environments. 

Following the automated system configuration, we partially automated the installation of the mitmproxy CA certificate but, the last step we make it manually. The user navigates to Settings/Security/Install from device storage/CA certificate and selects the certificate file that was previously transferred to the device's download directory. 

The actual traffic capture process begins when the user initiates the capture through the control script (the script interacts with the user through command line and it waits the user instruction to start capturing traffic). The system starts \texttt{tcpdump} with the command below to record all network traffic except communication between the device and mitmproxy proxy.

\begin{lstlisting}
tcpdump 
    -i any not port 8080 
    -w "/output/${NAME_APK}_${DAY_OF_CAPTURE}.pcap"
\end{lstlisting}

The process ID is stored for later termination control. Simultaneously, the \texttt{SSLKEYLOGFILE} environment variable ensures that TLS session keys are logged to \texttt{/output/sslkeylog.txt} for subsequent traffic decryption. During this phase, a human operator interacts with the installed app to generate realistic usage patterns, ensuring that captured traffic reflects authentic user behavior rather than automated testing scenarios. The capture process continues until the operator signals completion, at which point the system terminates \texttt{tcpdump} using the stored process ID and calculates the capture duration.

Upon capture completion, the system performs automated labeling operations to organize the collected traffic data. The PCAP file is moved from the temporary output directory to the permanent storage location with a standardized naming convention: \texttt{[name\_apk]\_[date]\_[duration].pcap}, where the filename includes the app name, timestamp, and capture duration in seconds. Similarly, the SSL key log file is processed and renamed as \texttt{sslkeylog\_[name\_apk]\_[date]\_[duration].txt} to maintain correlation between packet captures and decryption keys. This automated labeling system ensures that each capture session is uniquely identified and that corresponding PCAP and key files can be easily matched for analysis. The system also resets the SSL key log file by creating a new empty file for subsequent capture sessions, preventing key data from different sessions from being mixed.

Finally, the system prepares for the next capture session by restarting mitmproxy and optionally removing the test app. The mitmproxy service is restarted.
If the uninstall option was specified during script execution, the system removes the app using the following command to return the emulator to a clean state.
\begin{lstlisting}
./adb -s "\$EMULATOR\_ID" uninstall \$NAME\_APK
\end{lstlisting}
This reset process ensures that subsequent capture sessions start with a consistent environment, preventing interference between different app tests and maintaining the integrity of the traffic capture process. 
The automated reset capability enables researchers to conduct multiple capture sessions efficiently while maintaining experimental consistency across different apps or testing scenarios.

\subsection{System Limitations}

This subsection presents the limitations of our capture system related to mitmproxy traffic interception. These limitations affect DNS and QUIC traffic and app functionalities.

\textbf{DNS Protocol Limitations:} Our system configures the emulator with the \texttt{-dns-server 8.8.8.8} flag. When deploying the environment with or without mitmproxy, the resulting DNS traffic utilizes DoT queries despite this DNS server configuration. According to Android's cache resolver implementation, communications to 8.8.8.8 should utilize DoH protocol.

When operating without mitmproxy, DoH traffic becomes visible only when manually configuring Private DNS settings as \textit{dns.google} within the Android system. When mitmproxy is enabled, manual Private DNS configuration fails with \textit{Couldn't connect} error messages, and network communications stop working. System logs indicate SSL -1 errors when attempting to establish DoH connections.

\textbf{QUIC Protocol Limitations:} Without mitmproxy interference, all evaluated apps generate QUIC traffic, demonstrating widespread protocol adoption. However, mitmproxy deployment affects QUIC traffic visibility across apps. While some apps continue generating QUIC communications, others suppress QUIC traffic entirely when the mitmproxy is enabled.

\textbf{Certificate Pinning:} When capturing traffic with mitmproxy enabled, app functionality is limited due to certificate pinning security mechanisms. When apps detect unauthorized certificates, they stop working. We observed this limitation in all apps from the set of apps that we captured except Instagram, which continues to function with mitmproxy interception. For most apps, we only capture traffic during app launch, while for Instagram we capture both launch and normal traffic generation. When capturing traffic without mitmproxy interception, apps work normally.

Therefore, mitmproxy interception introduces limitations in app traffic generation. However, PARROT captures DoH and QUIC traffic generated by apps and their normal functionalities when operating without mitmproxy interception.

\section{Dataset Description}\label{sec:dataset_description}

To demonstrate our capture system capabilities and provide a traffic analysis dataset, we conducted a traffic collection campaign using the methodology described in Section~\ref{sec:capture_system}. This section presents the details of our dataset creation process, the apps studied, and the main characteristics of the collected traffic data.

\subsection{App Selection and Collection Campaign}

Our app selection process follows the MAppGraph dataset~\cite{Pham2021mappgraph}, which comprises 101 Android apps from the most popular apps in Vietnam in 2021. By using the same app set, we enable direct comparison between traffic datasets collected across different temporal periods. This comparative provides insights into the evolution of app traffic patterns between 2021 and the present.

Due to geographical restrictions and app availability limitations, we obtained APK files for 80 out of the 101 apps from the original MAppGraph dataset. These APK files were sourced from third-party repositories including Up2Down, APKMirror, and APKPure. For each app, we obtained the most recent version available that was compatible with our emulator configuration: \texttt{x86\_64} architecture and Android 15 (API level 35).

\begin{figure*}[!ht]
	\centering
	\includegraphics[width=\textwidth]{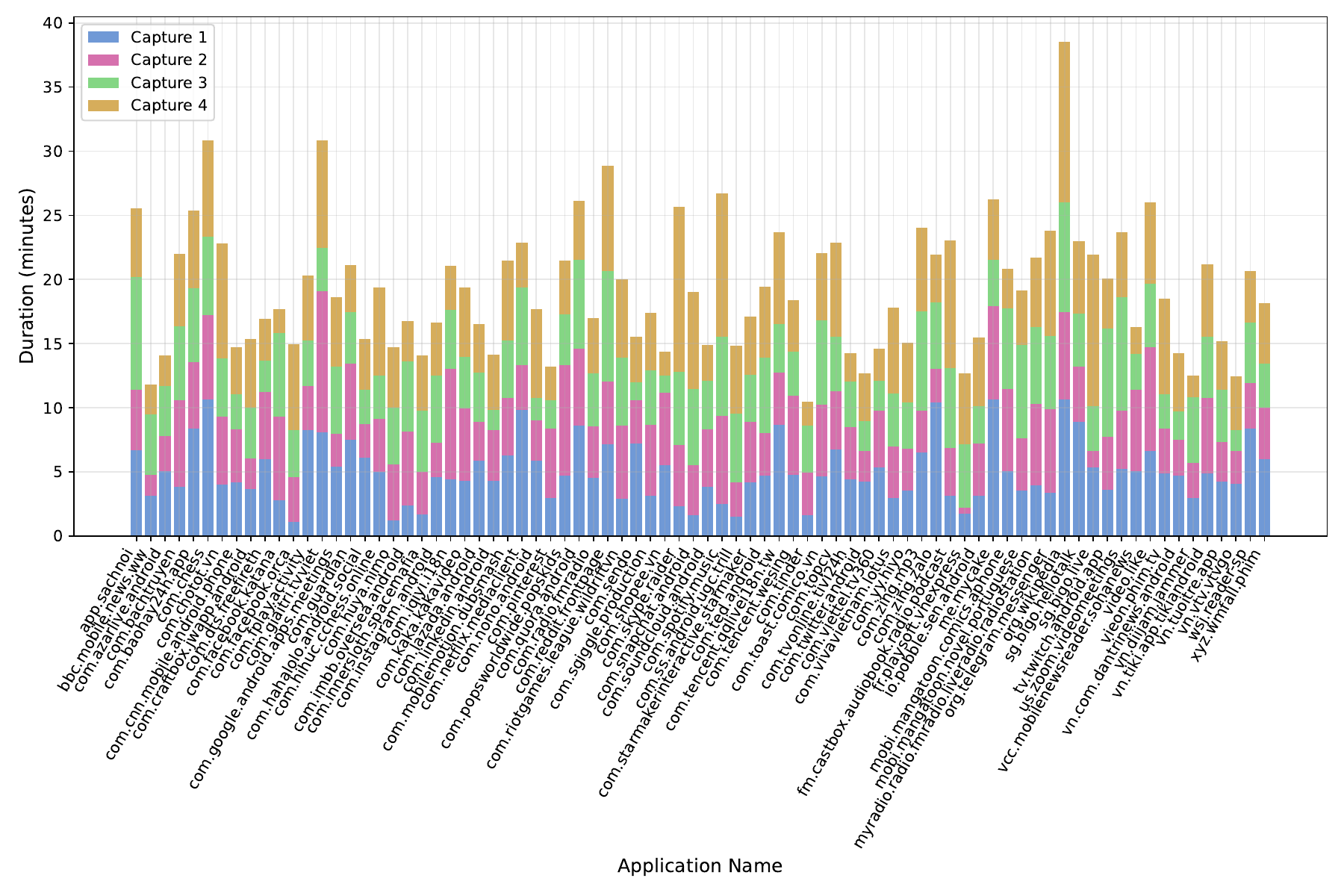}
	\caption{Apps captured and duration of traffic traces captured per app.}
	\label{fig:our-dataset_number-captures_duration}
\end{figure*}

Fig.~\ref{fig:our-dataset_number-captures_duration} presents an overview of our dataset, showing which apps where captured, the duration of each capture and the total traffic captured for each app. We captured 4 traffic traces for each of the 80 apps. The apps span categories includes social media, entertainment, news, gaming, communication, and e-commerce platforms.

The traffic collection campaign was conducted by members of our research group across multiple capture sessions from December 2024 to July 2025. For each app, we captured 4 traffic traces per app, with each trace lasting approximately 5 minutes. This approach resulted in a total of 320 traffic captures spanning more than 25 hours of app traffic. Our capture methodology focused on app initialization behavior, recording the traffic generated during the first deployment of each app following installation. 

\subsection{Traffic Interception}

During our campaign, we employed mitmproxy for traffic interception to obtain decryption keys for encrypted communications. Our system configuration uses Google Public DNS servers (8.8.8.8 and 8.8.4.4), which support encrypted DNS protocols by default, and mitmproxy is configured as a trusted Certificate Authority to enable TLS session key capture.

However, we encountered the challenge of certificate pinning, a security mechanism where apps ignore system-wide certificate authorities and trust only specific certificates predefined by the app developers. This mechanism substantially limited our capacity to decrypt app traffic. Among the 80 apps evaluated, only \texttt{Instagram} allowed login without connection errors and enabled full navigation through the app. Additionally, \texttt{Chess} successfully operated with mitmproxy enabled for offline game modes but presented limitations for online gameplay and login functionality. The remaining applications presented various types of limitations when mitmproxy intercepted traffic.

We provide a description of each app's behavior under mitmproxy interception in Appendix~\ref{appendix:app_behaviors}. The most frequent behaviors are: apps that require login to browse the app but we cannot login due to connectivity errors, such as \texttt{Skype} (com.skype.raider); apps that allow browsing without login but fail to load content and encounter connectivity errors during login attempts, such as \texttt{BBC} (bbc.mobile.news.ww); and apps that freeze during launch, such as \texttt{Messenger} (com.facebook.orca).

Despite these limitations, our dataset offers insights into app behavior under man-in-the-middle attack scenarios. Even when apps reject the mitmproxy certificate, they continue to produce observable network traffic patterns. Most importantly, DNS queries represent some of the first network activities as apps resolve domain names for their backend services during initialization. Although this DNS traffic uses encrypted versions (DoT), mitmproxy successfully provides the decryption keys for these communications, allowing us to identify which domains the apps are attempting to contact.

\subsection{Feature Extraction}

We developed a data processing pipeline that extracts network parameters from the raw PCAP files. Our processing methodology employs tshark, the command-line version of Wireshark, to extract structured data from packet captures. 

We handle both IPv4 and IPv6 traffic across TCP and UDP transport protocols. For each packet, we extract the following parameters:  the timestamp, the IP source and IP destination address, the source and destination port, the transport protocol, the app protocol if exists, the additional information of the packet and the packet length.

We gave TLS traffic special treatment. To ensure accurate extraction of TLS packet versions, we extract version information from two sources: the TLS handshake version field (\texttt{tls.handshake.version}) and the supported versions extension (\texttt{tls.handshake.extensions.supported\_version}).

\subsection{Dataset Characteristics}\label{Dataset_Characteristics}

The resulting dataset provides a view of app traffic patterns captured in a controlled environment. All traffic originates from the emulator IP address (172.29.184.165), providing clean app labeling and eliminating background traffic contamination that complicates real-world traffic analysis. The controlled environment ensures that all captured traffic is attributable to the specific app under test.

\begin{figure}[!h]
    \centering
    \subfigure[Communication graph of a subset of 10 apps of our dataset.]{
        \includegraphics[width=\columnwidth]{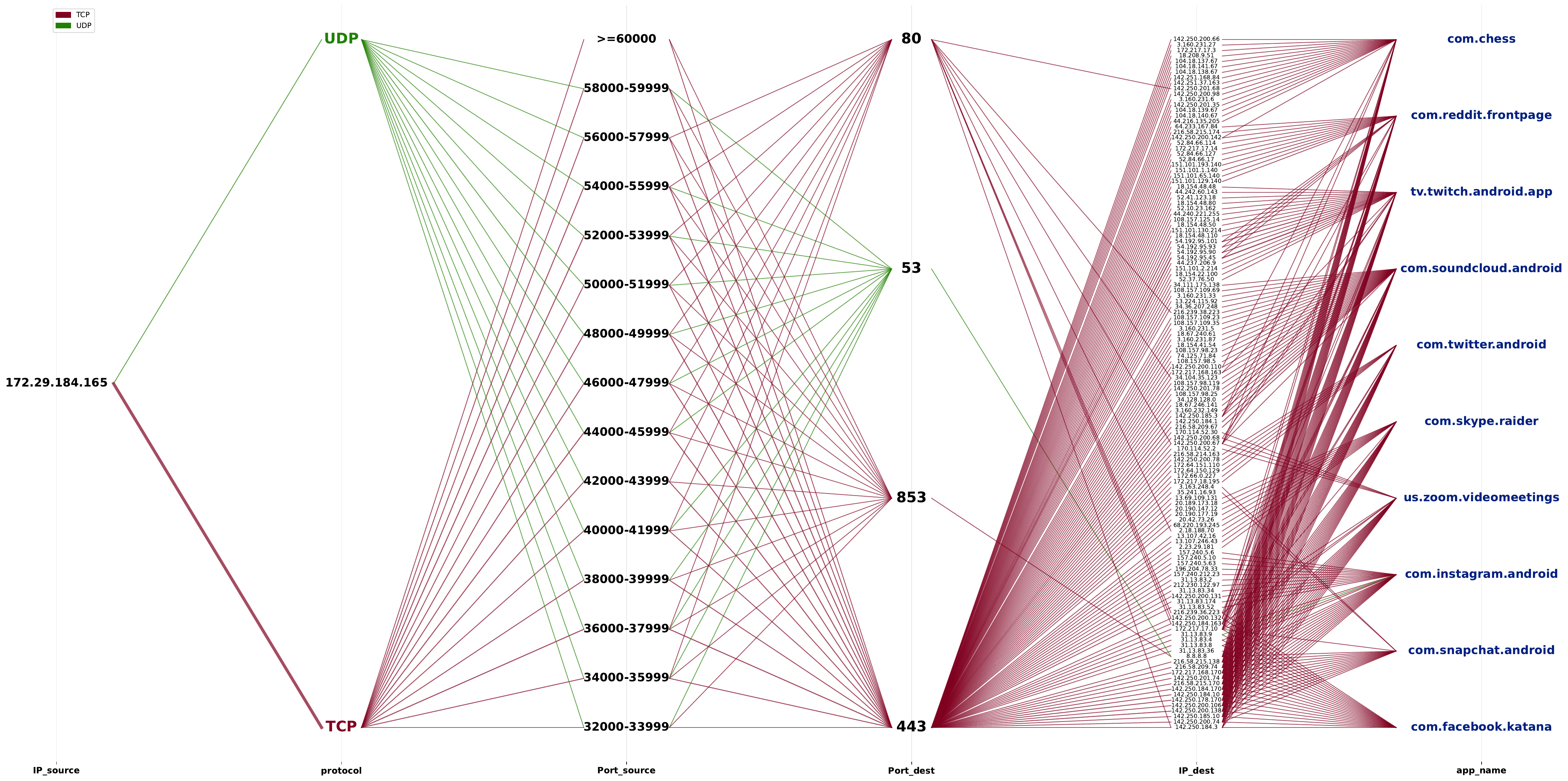}         \label{fig:network_communication_graph_subset-10apps}
    }
    \hfill
    \subfigure[Communication graph of the app Reddit]{
        \includegraphics[width=\columnwidth]{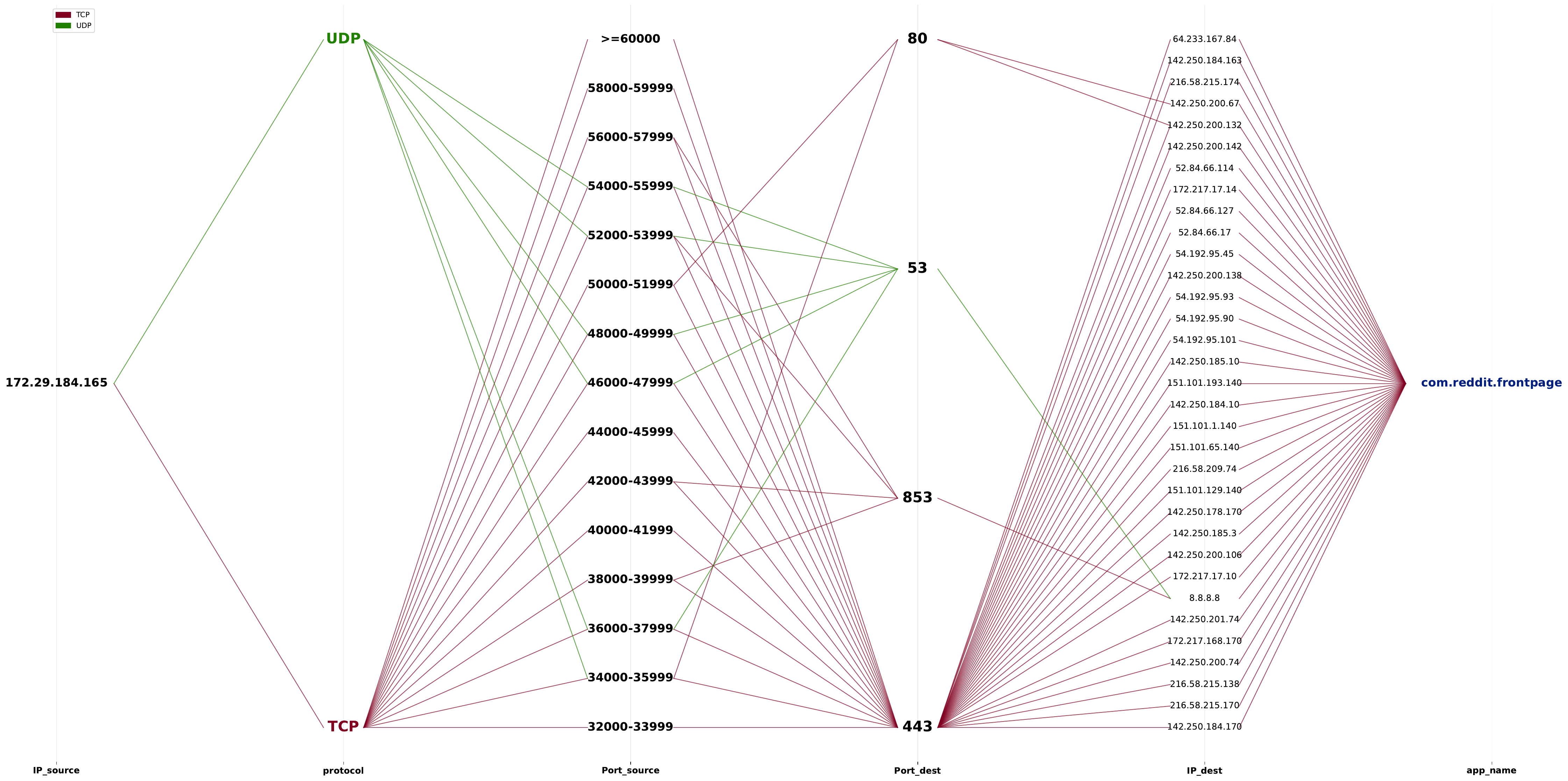}         \label{fig:network_communication_graph_reddit}
    }
    \caption{Communication graphs that represent the traffic flow generated by the apps.}
    \label{fig:comm_graphs}
\end{figure}

Fig.~\ref{fig:comm_graphs} presents communication graph visualizations of our dataset traffic patterns. Fig.~\ref{fig:network_communication_graph_subset-10apps} shows traffic from 10 apps, representing a subset of our dataset, while Fig.~\ref{fig:network_communication_graph_reddit} displays the specific traffic flow patterns of the Reddit app. Both graphs display six columns representing the flow of network communications: IP source, transport protocol, source port intervals, destination ports, destination IP addresses, and app names. Each connection between columns represents actual packet flows observed in our captures. 

The source port intervals are grouped for better visualization, as apps typically use ephemeral ports for outgoing connections. The destination ports reveal distinct communication patterns: port 443 dominates app traffic representing HTTPS communications, port 853 is used for \ac{DoT} traffic, while ports 80 and 53 appear for connectivity checks.

Traffic on ports 80 and 53 occurs due to Android's captive portal detection mechanism~\cite{android_captive_portal}. 
Apps perform connectivity checks by contacting specific URLs (such as \texttt{connectivitycheck.gstatic.com}) over HTTP port 80 to verify internet access and detect network restrictions. The DNS queries over port 53 serve the same connectivity verification purpose, and these queries typically follow the HTTP connectivity checks. 

We highlight several destination IP addresses in our dataset. The IP address 8.8.8.8, belonging to Google's Public DNS service, receives both port 53 and port 853 traffic for DNS resolution. Furthermore, the top 10 most frequently contacted IP addresses across all apps include: 54.192.95.45 (Amazon CloudFront), 142.250.184.3 and 142.250.200.138 (Google services), along with several other Google and content delivery network addresses. These patterns demonstrate the modern app ecosystem's dependence on cloud services, content delivery networks, and Google's infrastructure.

\subsection{Background Traffic Analysis}

Our controlled environment enables precise app labeling since only the target app and system components are active during capture sessions. There is only one app installed apart from the apps installed by default, and there is no residual traffic from other apps or apps running in the background. To characterize the baseline traffic generated by the Android system without user-installed apps, we captured traffic in our scenario with mimtproxy enabled from a clean emulator instance with no additional apps installed beyond the default system apps.

Table~\ref{tab:background_traffic_summary} presents the packet distribution across transport and app protocols for the background traffic capture session. We observed that we mostly have TCP encrypted traffic but also found unencrypted protocols: HTTP packets and Do53 packets as well as the captures with traffic generated by apps. The HTTP traffic (port 80) consists of connectivity check requests to \texttt{connectivitycheck.gstatic.com/generate\_204}. The Do53 traffic also makes queries for connectivity checks to the domains \texttt{connectivitycheck.gstatic.com} and \texttt{www.google.com}. The TCP encrypted traffic uses TLSv1.3 protocol and we also found DoT traffic that generates queries to Google services domains: \texttt{gmscompliance-pa.googleapis.com}, \texttt{android.googleapis.com}, and \texttt{digitalassetlinks.googleapis.com}.

\begin{table}[ht]
\centering
\caption{Background traffic protocol distribution}
\label{tab:background_traffic_summary}
\begin{tabular}{lcc}
\hline
Transport Protocol & App Protocol & Packet Count \\
\hline
UDP & DNS & 14 \\
TCP & HTTP & 4 \\
TCP & TLSv1.3 & 489 \\
TCP & DoT & 19 \\
\hline
\end{tabular}
\end{table}

\begin{figure}[!h]
    \centering
    \includegraphics[width=\columnwidth]{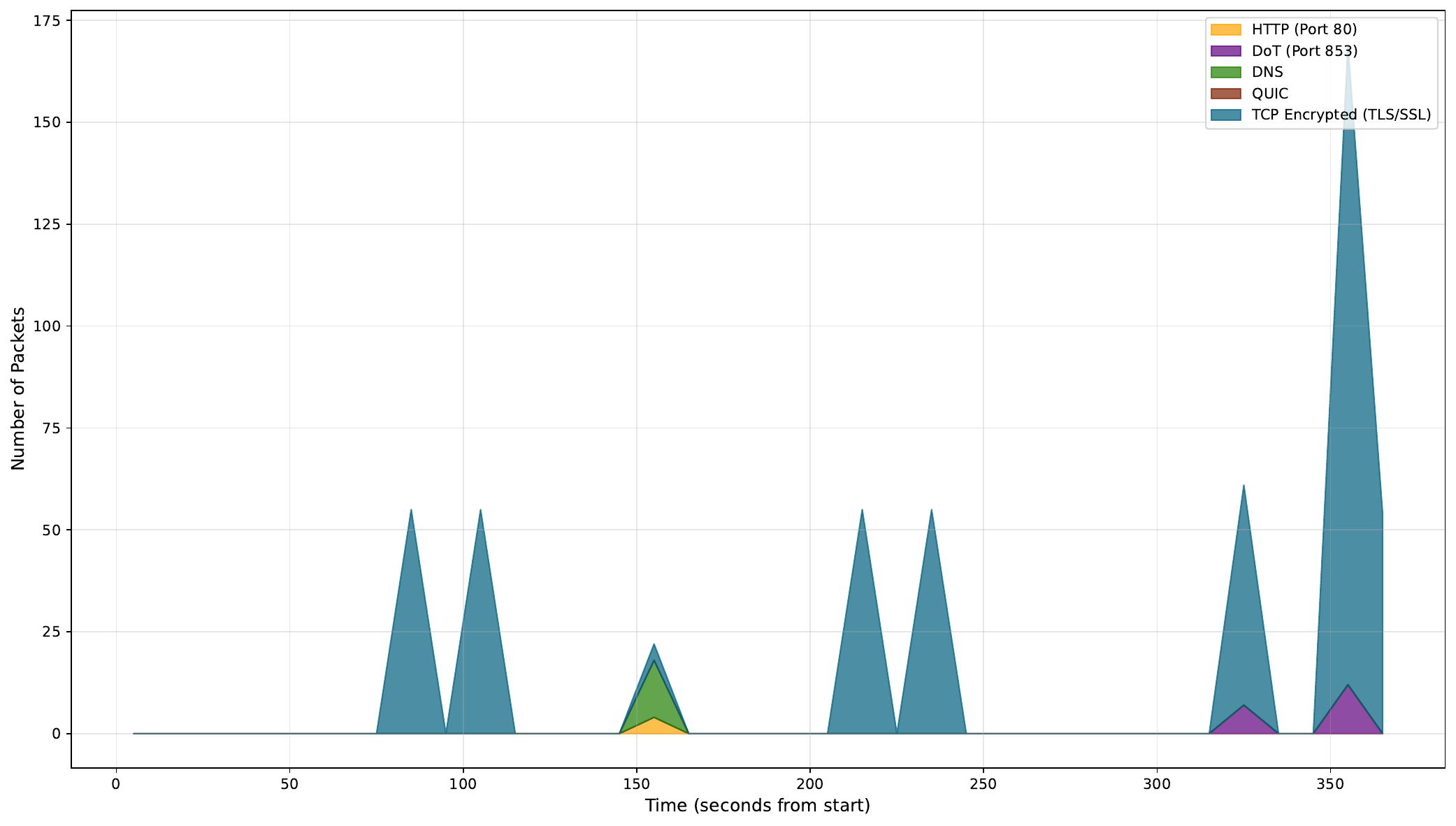}
    \caption{Background traffic pattern generated by Android system without user-installed apps. The stacked area plot shows the temporal distribution of different protocols over a 5-minute period.}
    \label{fig:background_traffic}
\end{figure}

Fig.~\ref{fig:background_traffic} presents the temporal distribution of background traffic generated by the Android system over a 5-minute capture period. We only represent the app data packets. The stacked area plot displays packet counts binned into 10-second intervals across five protocols: HTTP traffic (orange), DoT (purple), standard DNS queries (green), QUIC protocol (dark red), and TCP encrypted traffic (blue).

The network pattern shows that TCP encrypted traffic represents the majority of background communications. HTTP and Do53 traffic related to connectivity checks occur during the initial period of the capture, while DoT traffic appears at the end of the capture session. The background traffic analysis reveals characteristics of Android's security mechanisms when routing TCP traffic through mitmproxy. This demonstrates that our app-specific captures contain minimal system-generated traffic. We can therefore attribute any HTTP and Do53 traffic observed in app captures to Android's default connectivity verification mechanisms rather than app-specific behavior. This low baseline ensures that captured traffic is predominantly attributable to the target app, providing clean app labeling for analysis purposes.

\section{Comparative Analysis of App Traffic Evolution}\label{sec:analysis_datasets}

In this section, we present a comparative analysis between our dataset and the MAppGraph dataset to examine the evolution of app traffic patterns from 2021 to 2025. Due to the temporal and geographical differences in dataset capture locations (MAppGraph 2021 in Vietnam and our dataset 2025 in Spain), we obtained 50 common apps from the 80 apps that the authors share with the research community. For fair comparison, we consider only the first 5 minutes of the MAppGraph captures since our captures have a duration of approximately 5 minutes. Our analysis focuses on protocol distribution, traffic metrics, encryption trends, and DNS protocol evolution across these 50 common apps.

\subsection{Protocol Overview and Traffic Distribution}\label{Protocol_Overview}

\begin{figure}[!h]
    \centering
    \subfigure[Protocol usage overview of our dataset.]{
        \includegraphics[width=0.48\columnwidth,clip,trim=2.2cm 2cm 6.5cm 2cm]{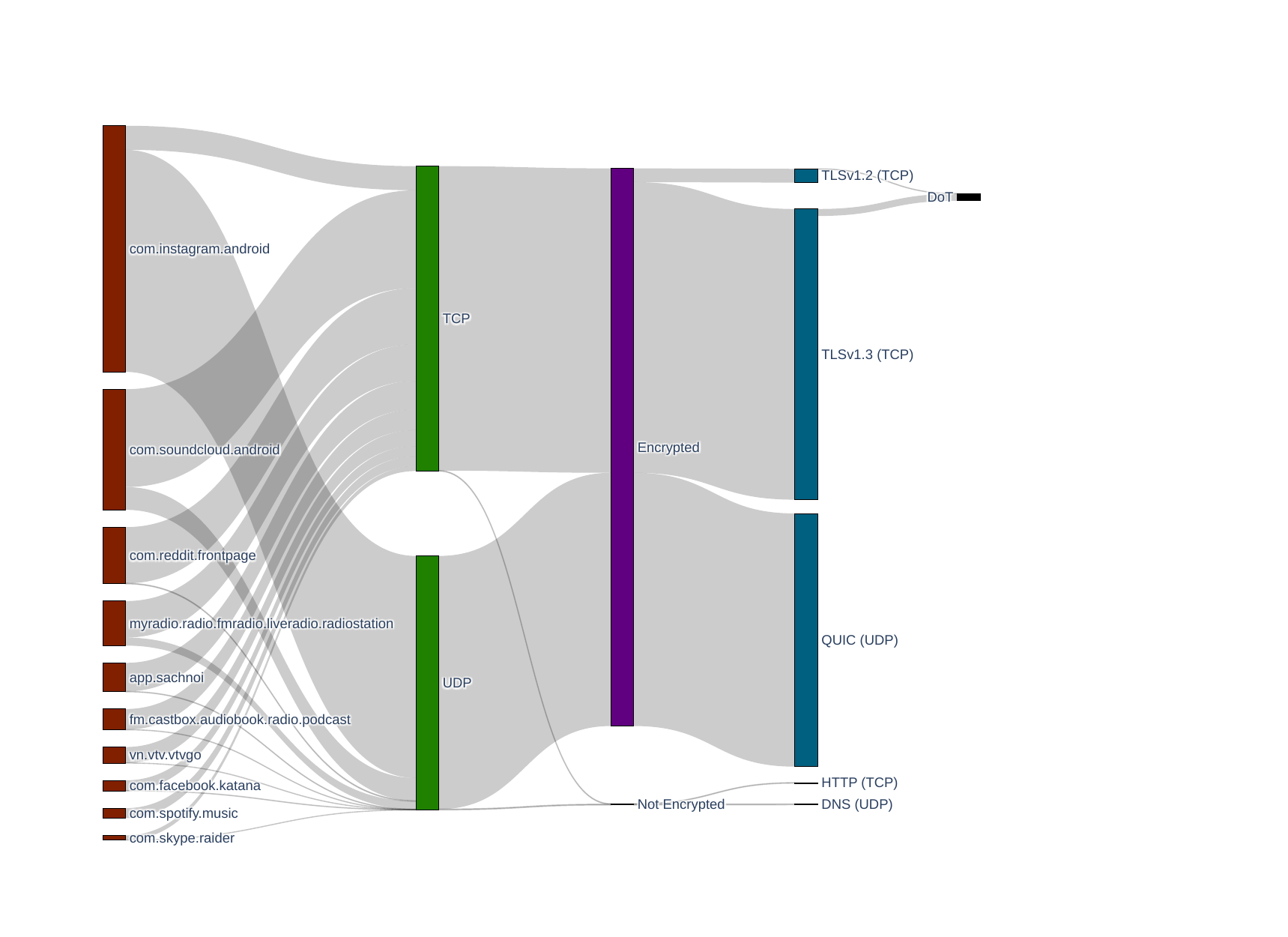}
        \label{fig:sankey_our}
    }
    \subfigure[Protocol usage overview of MAppGraph dataset.]{
        \includegraphics[width=0.48\columnwidth,clip,trim=2.2cm 2cm 6.5cm 2cm]{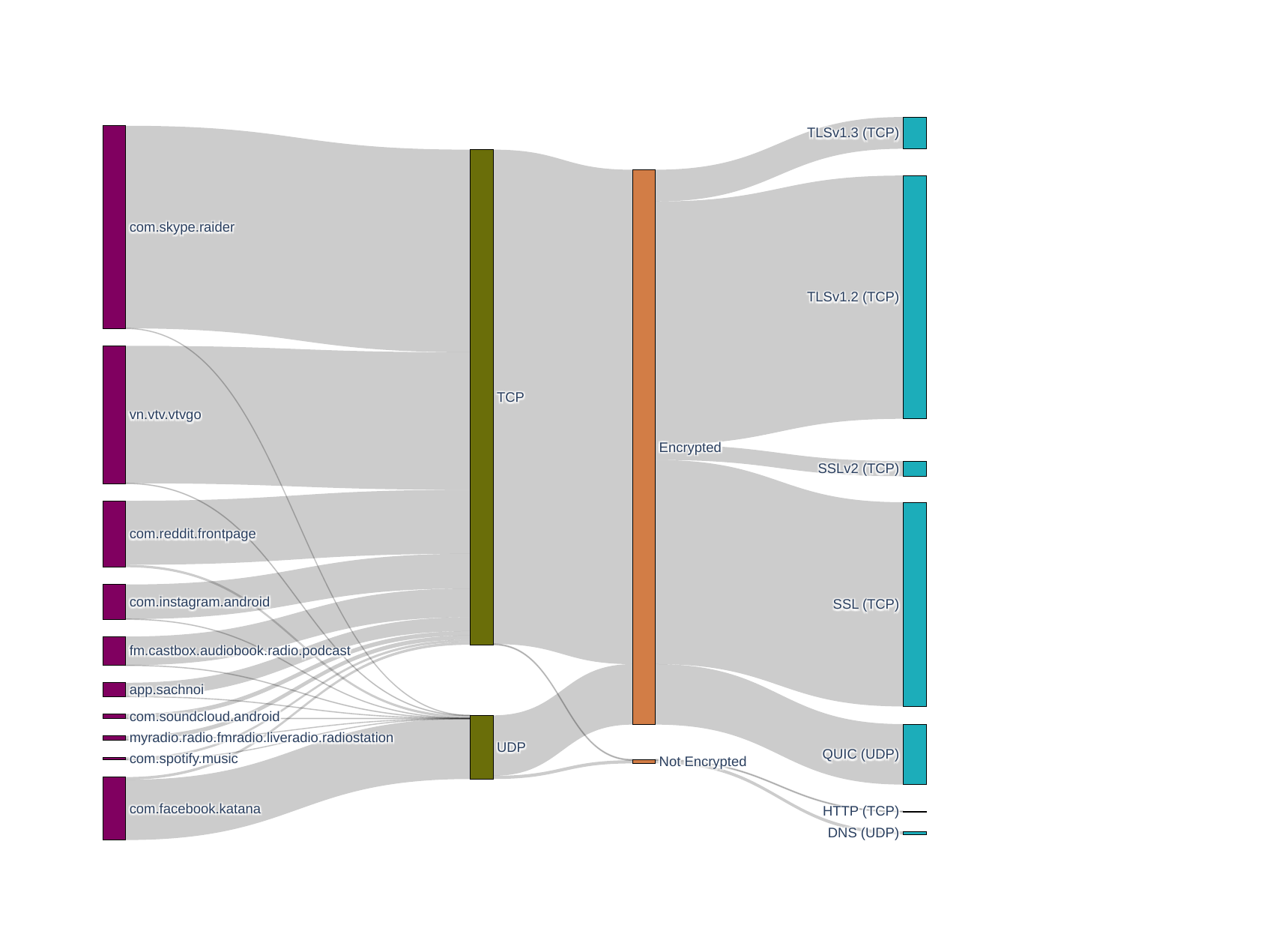}
        \label{fig:sankey_mappgraph}
    }
    \caption{Sankey diagrams showing protocol usage overview across 10 common apps of both datasets. The flow visualization illustrates the relationship between transport protocols, the status of the encryption and app protocols for each app in (a) our captured dataset and (b) the MAppGraph dataset.}
    \label{fig:sankey_comparison}
\end{figure}

Fig.~\ref{fig:sankey_comparison} presents Sankey diagrams that visualize protocol distribution across both datasets for a subset of 10 common apps. These diagrams illustrate the flow from transport protocols (TCP/UDP) through encryption status to app protocols, representing only packets containing app data. The apps are the same in both diagrams but are sorted by traffic volume within each dataset, which varies depending on the specific traffic execution characteristics of each capture session.

The encrypted protocol distribution reveals the evolution of app communications between 2021 and 2025. Transport protocol usage shows our dataset with a more balanced distribution of TCP (54.6\%) and UDP (45.4\%) compared to MAppGraph's TCP-dominated traffic (88.6\% TCP vs. 11.4\% UDP). This shift reflects the increased adoption of QUIC protocol, which operates over UDP and accounts for 45.3\% of our dataset compared to only 10.8\% in MAppGraph.

The encryption landscape demonstrates a migration toward more secure protocols. Our 2025 dataset shows TLSv1.3 as the dominant encrypted protocol at 52.0\% with minimal TLSv1.2 usage at 2.5\%. In contrast, the MAppGraph dataset from 2021 relies heavily on older encryption protocols: TLSv1.2 dominating at 43.5\%, SSL accounting for 36.5\%, and TLSv1.3 representing only 5.7\%. This evolution indicates the industry-wide migration toward more secure and efficient TLS versions, with TLSv1.3 becoming the standard. 

The emergence of \ac{DoT} protocol appears in our dataset. We identified 1.3\% DoT traffic of total traffic, with TLSv1.3 comprising 97.7\% of these encrypted DNS communications. The MAppGraph dataset contains no DoT traffic, even though built-in support for encrypted DNS has been included since Android 9~\cite{android_dns_over_tls_blog}, 
where the device automatically upgrades to DoT traffic but this functionality depended on the DNS server support. Therefore, the evolution of DNS traffic demonstrates a security enhancement as DoT traffic support has become widely extended. In Subsection~\ref{dns_evolution} we provide more detail about this evolution.

These protocol evolution patterns indicate a app ecosystem that has enhanced security through encrypted DNS adoption, improved performance through increased QUIC utilization, and achieved a transition from older TLS versions to TLSv1.3. The changes reflect technological advancement and security awareness in app development between both datasets captured in 2021 and 2025.

\subsection{Packets Per Minute And Intensity of Traffic}

To provide context for our comparative analysis, we examine the packet transmission rates of each app, as traffic volume can vary significantly due to capture execution characteristics and traffic collection scenarios. Our dataset focuses exclusively on app deployment behavior, capturing traffic generated during the initial startup phase following installation. In contrast, the MAppGraph dataset, despite considering only the first 5 minutes of each capture, includes traffic from active user navigation within the apps. This methodological difference is reflected in the observed traffic volume variations between the datasets.

\subsubsection{Packets per minute}
Table~\ref{tab:ppm_comparison_10apps} presents mean packet rate comparisons between our dataset and the MAppGraph dataset for the same 10 apps analyzed in the protocol distribution shown in Fig.~\ref{fig:sankey_comparison}. The packet rates represent the average packets per minute calculated across all capture sessions for each app in both datasets. The data illustrates the variability in traffic patterns across different app categories.

\begin{table}[ht]
\centering
\caption{Packet rate comparison for selected apps (packets per minute)}
\label{tab:ppm_comparison_10apps}
\begin{tabular}{lcc}
\hline
APK Name (App) & Our Dataset & MAppGraph \\
\hline
app.sachnoi (Sachnoi) & 2,278 & 5,626 \\
com.facebook.katana (Facebook) & 20,310 & 6,385 \\
com.instagram.android (Instagram) & 8,081 & 10,547 \\
com.reddit.frontpage (Reddit) & 4,047 & 18,142 \\
com.skype.raider (Skype) & 1,588 & 71,584 \\
com.soundcloud.android (SoundCloud) & 7,988 & 3,424 \\
com.spotify.music (Spotify) & 2,248 & 2,443 \\
fm.castbox.audiobook.radio.podcast (Podcast Player) & 3,166 & 14,728 \\
myradio.radio.fmradio.liveradio.radiostation (My Radio) & 3,583 & 24,067 \\
vn.vtv.vtvgo (VTV Go) & 4,516 & 50,444 \\
\hline
\end{tabular}
\end{table}

The packet transmission rates show that the MAppGraph dataset has higher overall traffic intensity compared to our 2025 captures, as expected a priori. Across 50 common apps, MAppGraph exhibits an average packet rate of 21,288 packets per minute (ppm), while our dataset averages 4,019 ppm, representing a 5.3x difference in traffic volume. Our dataset exhibits higher packet rates for 8 apps: Chess (7.53x higher), Facebook (3.18x higher), Facebook Messenger (3.15x higher), SoundCloud (2.33x higher), Among Us (1.83x higher), Likee (1.78x higher), Noveltoon (1.45x higher), and StarMaker (1.31x higher). These apps show more traffic in our captures, which may be due to less navigation during the initial minutes of MAppGraph captures or more network queries during the initial deployment phase of these apps.

\subsubsection{Temporal Traffic Distribution Patterns}

After illustrating the differences in protocol usage and mean packet rates between datasets, we now examine the traffic distribution patterns of specific app captures. The temporal distribution of app traffic reveals distinct patterns when examining individual capture sessions through stacked area visualizations. These plots represent app data packets binned into 10-second intervals, displaying four protocol categories: TCP encrypted traffic (TLS/SSL) in blue, QUIC protocol in dark red, standard DNS queries in green, and \ac{DoT} traffic in purple.

\begin{figure*}[!ht]
    \centering
    \subfigure[Reddit - Our Dataset (2025)]{
        \includegraphics[width=0.48\textwidth]{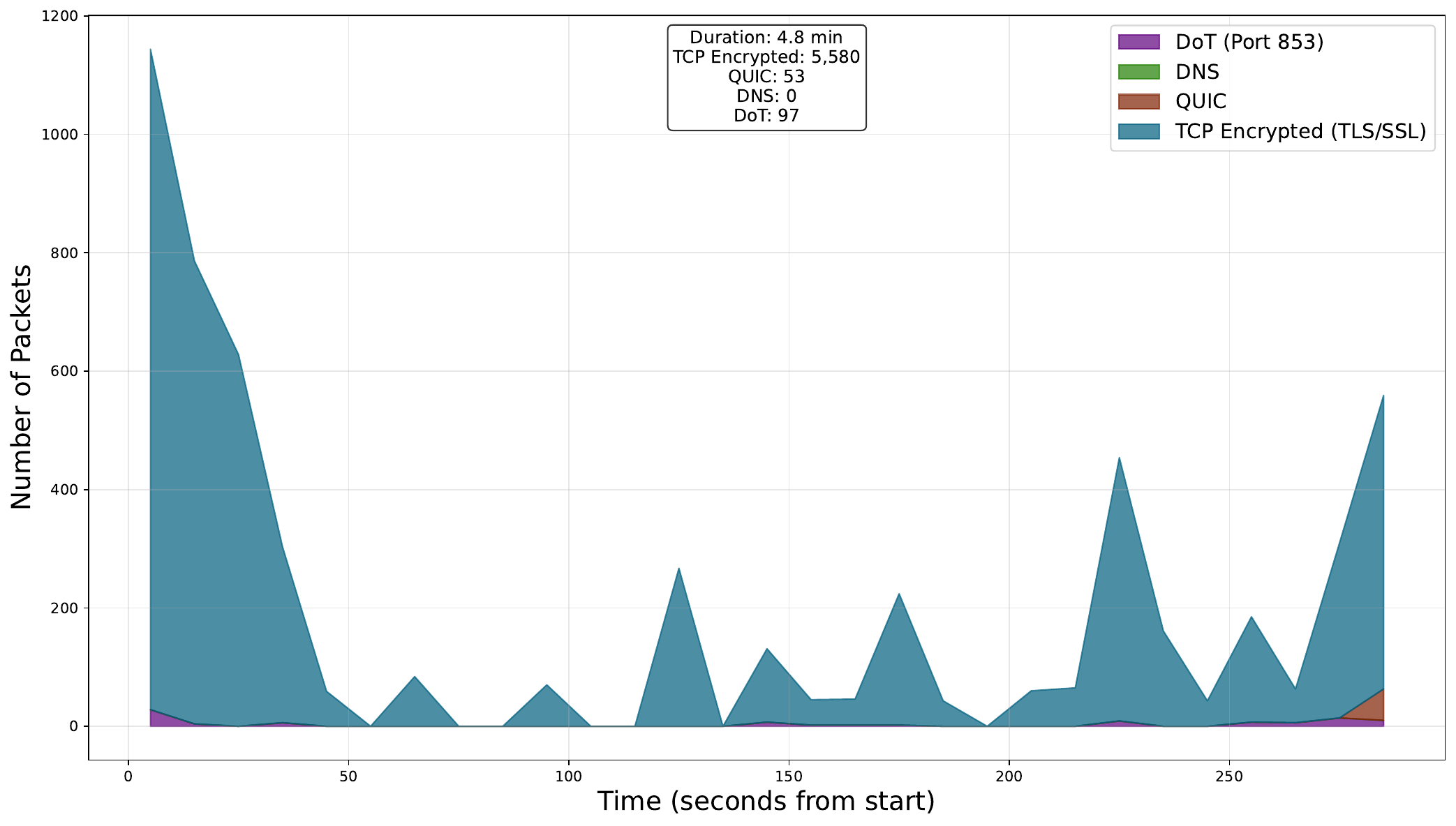}
        \label{fig:reddit_stacked_area_our}
    }
    \subfigure[Reddit - MAppGraph Dataset (2021)]{
        \includegraphics[width=0.48\textwidth]{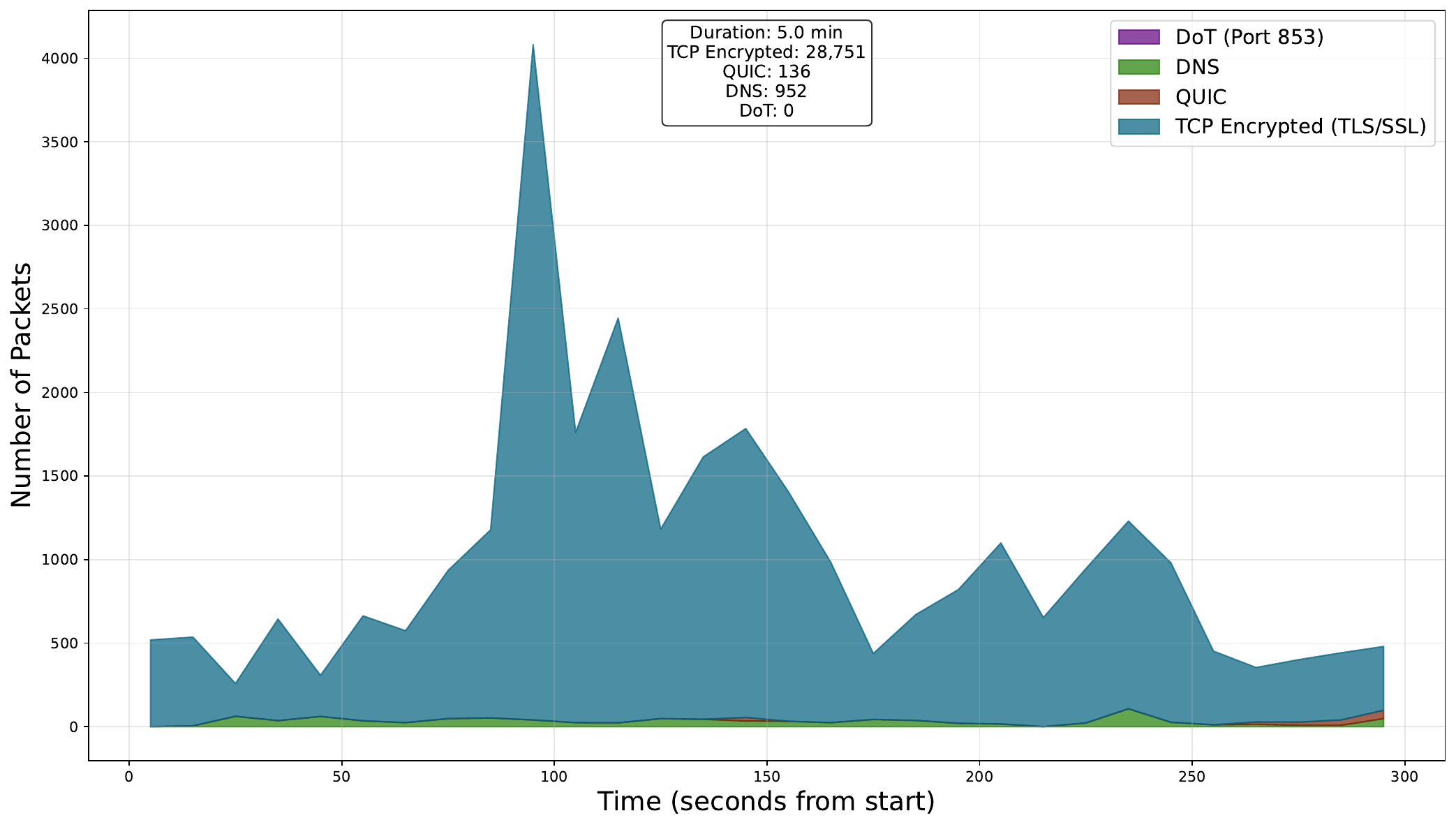}
        \label{fig:reddit_stacked_area_mappgraph}
    }
    \subfigure[SoundCloud - Our Dataset (2025)]{
        \includegraphics[width=0.48\textwidth]{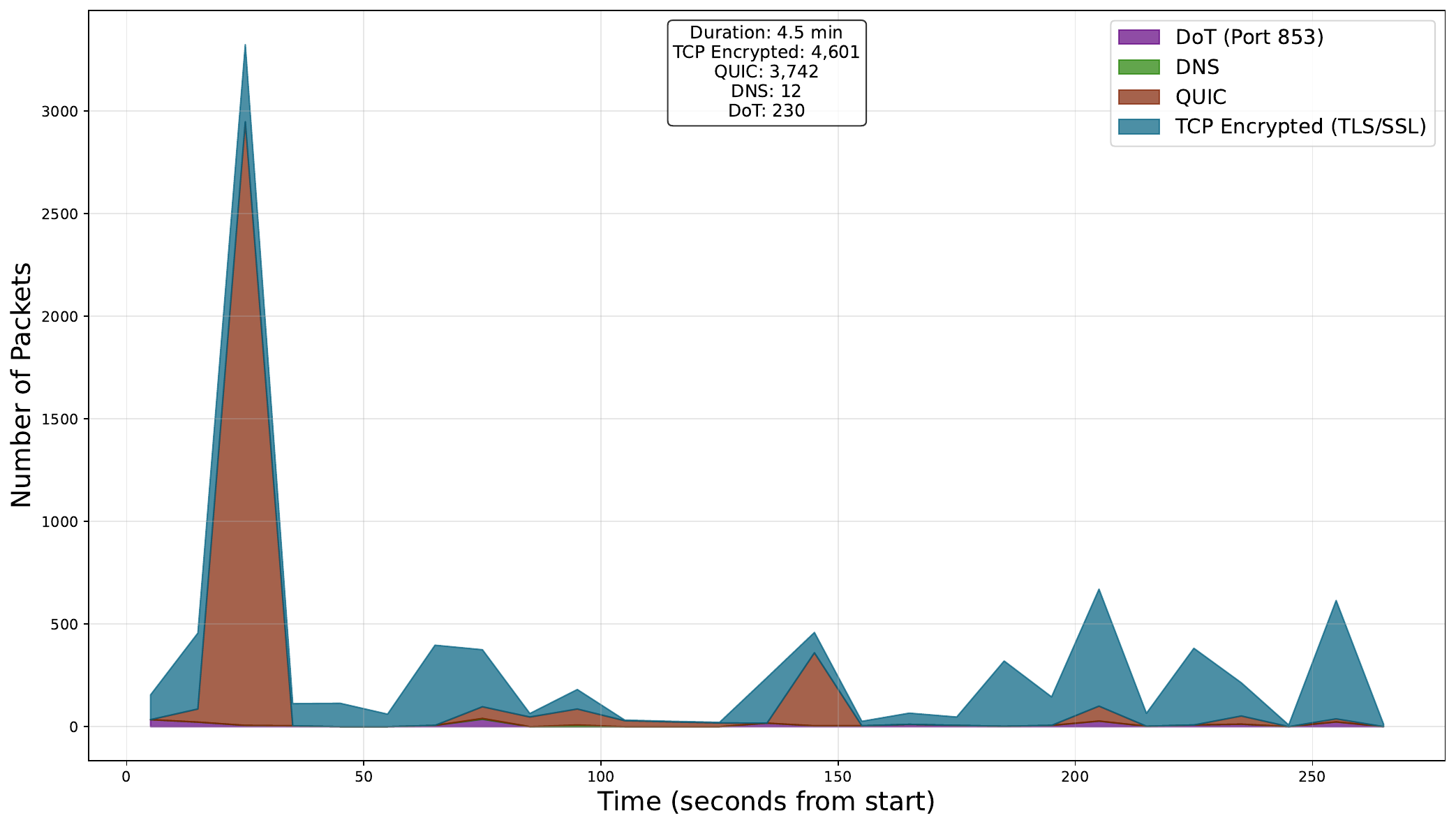}
        \label{fig:soundcloud_stacked_area_our}
    }
    \subfigure[SoundCloud - MAppGraph Dataset (2021)]{
        \includegraphics[width=0.48\textwidth]{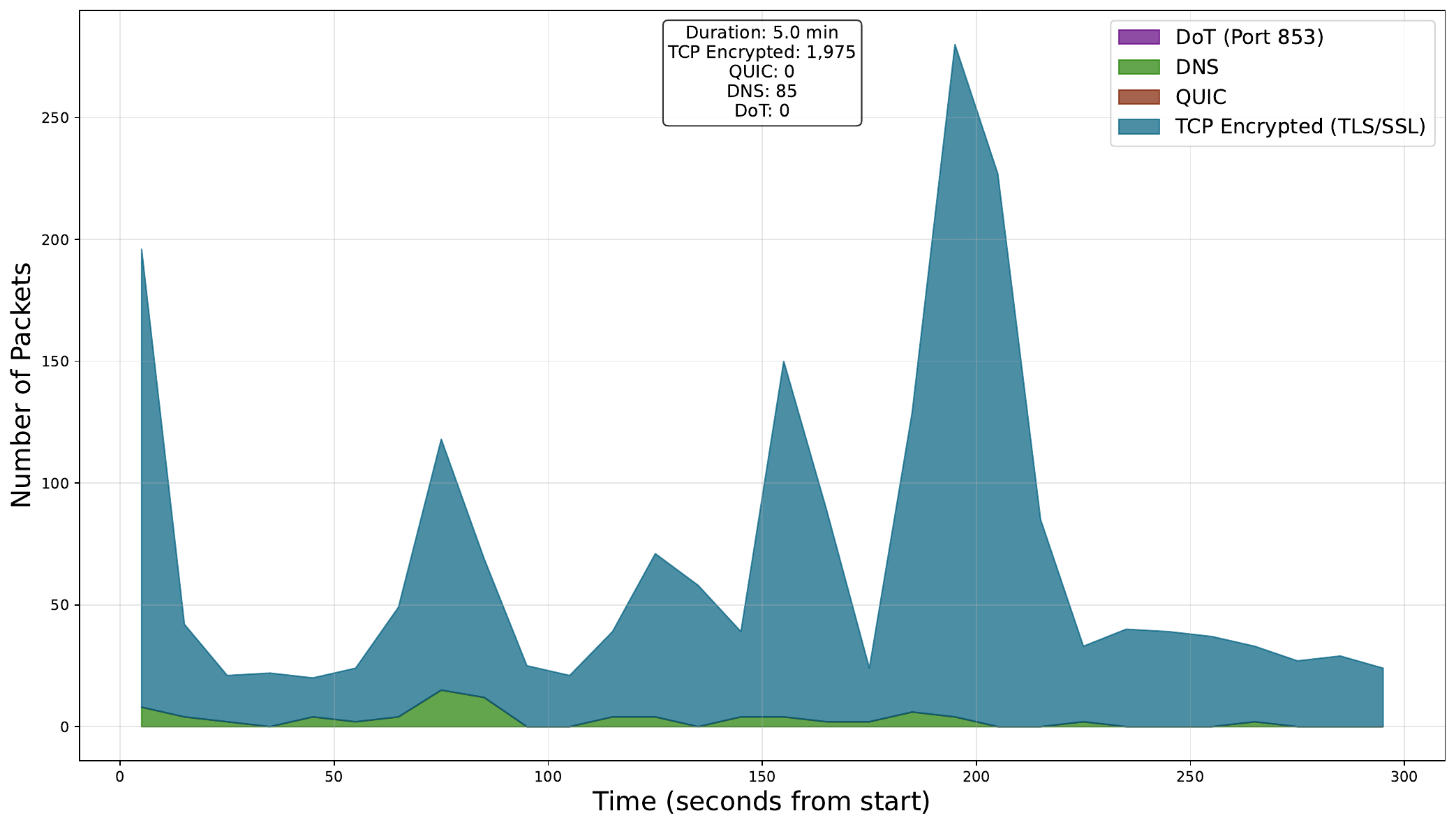}
        \label{fig:soundcloud_stacked_area_mappgraph}
    }
    \caption{Temporal traffic distribution patterns showing stacked area plots for Reddit and SoundCloud apps. Each plot displays app data packets binned into 10-second intervals across four protocol categories: TCP encrypted traffic (blue), QUIC protocol (dark red), DNS queries (green), and DoT (purple). The comparison illustrates traffic intensity patterns between traffic captures of MAppGraph dataset and our dataset.}
    \label{fig:stacked_area_comparison}
\end{figure*}

Fig.~\ref{fig:stacked_area_comparison} illustrates traffic patterns of the app data packets for Reddit and SoundCloud apps across both datasets. The visualization demonstrates how different capture scenarios and execution characteristics affect the traffic patterns generated by apps. Additionally, the comparison reveals the evolution of protocol usage between 2021 and 2025 captures. Each plot includes statistical summaries showing total packets, capture duration, and packet counts for each protocol.

Reddit traffic patterns show distinct initialization phases in both datasets, with our 2025 captures exhibiting more concentrated traffic bursts during the initial 50 seconds, followed by periods of reduced activity denoting the effect of the Android security mechanism against mimtproxy scenario. The MAppGraph Reddit capture demonstrates more sustained traffic throughout the entire session, with consistent packet rates across multiple time intervals. Encrypted TCP traffic dominates both captures, with some QUIC traffic usage appearing at the end of the 5-minute capture session. DNS traffic is present throughout both captures: our dataset shows DoT traffic with 97 packets (including requests and responses), while the MAppGraph capture uses Do53 traffic with more intensive usage totaling 952 packets (including requests and responses).

SoundCloud traffic exhibits similar temporal characteristics, with our dataset showing concentrated traffic bursts during the initial seconds followed by periods of reduced activity, while the MAppGraph capture presents more sustained traffic throughout the entire session. However, the protocol usage differs significantly between datasets. Our capture demonstrates substantial QUIC protocol usage, reflecting the adoption of modern encrypted protocol versions, in contrast to the MAppGraph capture which presents a traditional TCP-dominated pattern. DNS traffic appears throughout both captures, with our dataset showing more DNS packets using the encrypted DoT version compared to the MAppGraph capture, indicating different traffic execution characteristics.

These temporal patterns provide insights into the traffic characteristics of each dataset. The analysis reveals the effect of Android security mechanisms when using mitmproxy, which results in lower traffic rates. Nevertheless, the analysis demonstrates a clear migration from traditional TCP-based encrypted communications to QUIC protocol and the adoption of encrypted DNS traffic (DoT), reflecting enhanced security implementations in apps between 2021 and 2025. These findings demonstrate the importance of considering both temporal evolution and methodological factors when comparing app traffic datasets across different collection periods.

\subsection{Encryption Protocol Evolution}\label{encryption_evolution}

In this subsection, we provide insights into the evolution of encrypted protocols specifically. The analysis reveals significant changes in app security implementations between 2021 and 2025. We examined TCP encrypted protocols (SSL/SSLv2/TLSv1/TLSv1.2/TLSv1.3) and QUIC as the encrypted UDP protocol across the common apps present in both datasets.

\begin{figure}[htbp]
    \centering
    \subfigure[TCP encrypted protocols]{
        \includegraphics[width=0.48\columnwidth]{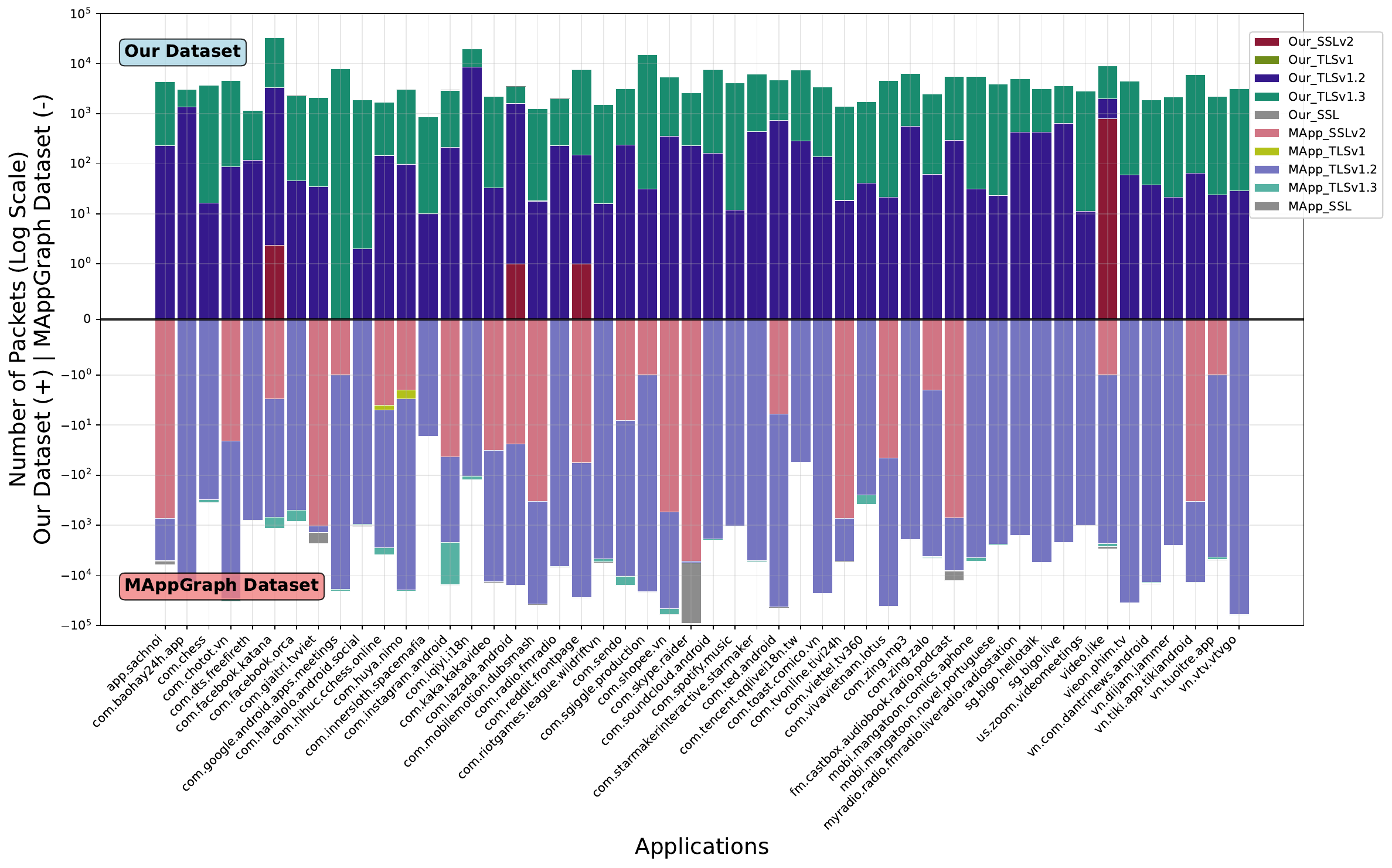}
        \label{fig:TCP_bihistogram_ourdatasetVSmappgraph}
    }
    \subfigure[QUIC protocol]{
        \includegraphics[width=0.48\columnwidth]{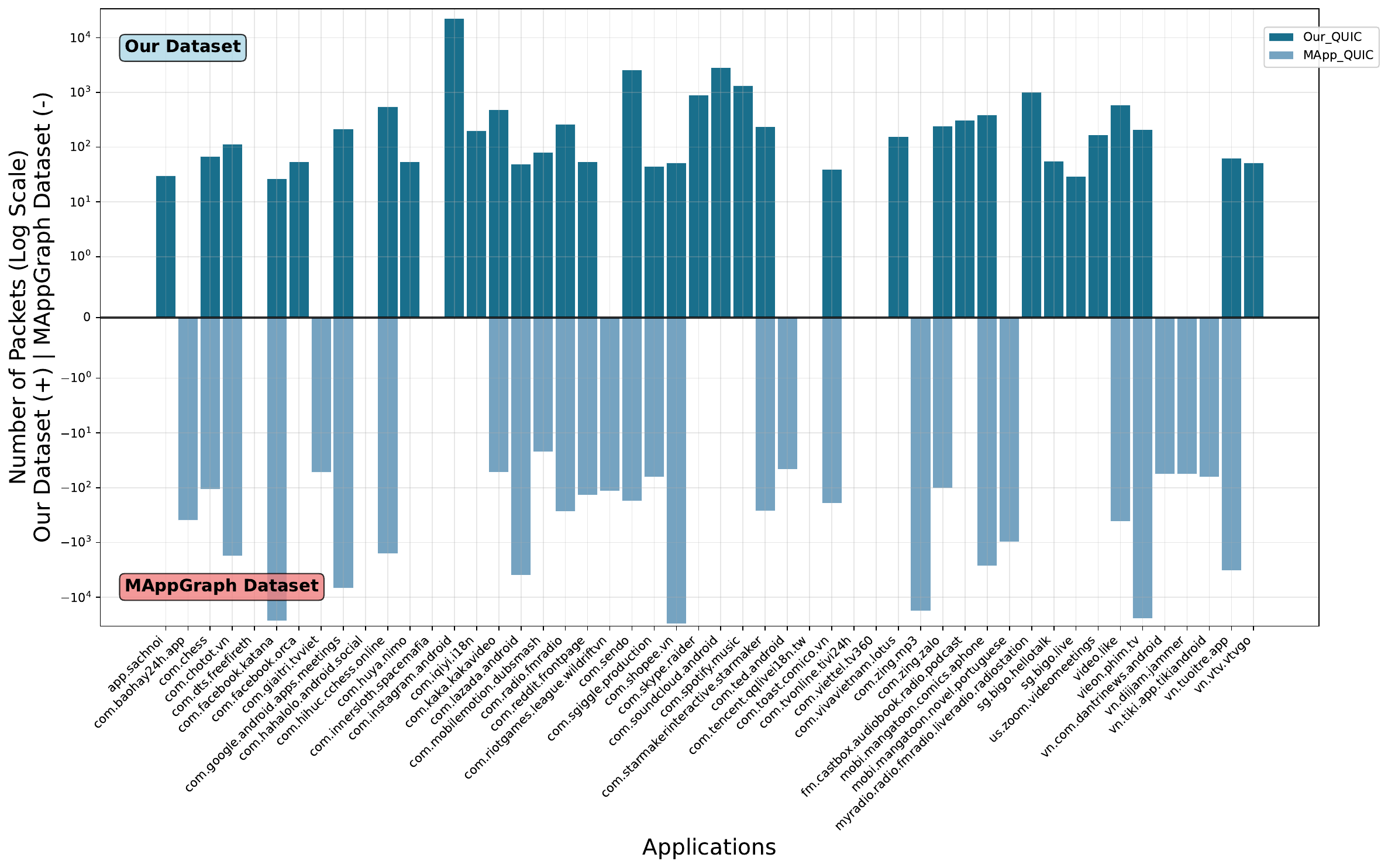}
        \label{fig:UDP_bihistogram_ourdatasetVSmappgraph}
    }
    \caption{Bihistograms comparing encrypted protocol distribution between our dataset (positive bars) and MAppGraph dataset (negative bars). (a) TCP encrypted protocols showing the migration from legacy SSL/TLS versions to TLSv1.3. (b) QUIC protocol adoption patterns across apps.}
    \label{fig:bihistogram_encrypted_protocols}
\end{figure}

Fig.~\ref{fig:bihistogram_encrypted_protocols} presents bihistograms comparing encrypted protocol usage between our dataset (positive bars) and the MAppGraph dataset (negative bars) for both TCP and UDP protocols. The TCP bihistogram displays five protocols: SSLv2 (burgundy/pink), TLSv1 (olive/yellow), TLSv1.2 (navy/light blue), TLSv1.3 (teal/green), and SSL (grey), while the UDP bihistogram focuses on QUIC protocol usage (dark blue/light blue). We used a logarithmic scale to accommodate the wide range of packet counts across apps.

The TCP encrypted protocol landscape demonstrates a substantial migration from older encryption standards to modern TLS implementations. Our 2025 dataset shows TLSv1.3 comprising 90.0\% of all TCP encrypted traffic represented in teal coloring in the positive bars in Fig.~\ref{fig:TCP_bihistogram_ourdatasetVSmappgraph}, compared to only 6.7\% in the MAppGraph 2021 dataset represented in green in the negative bars. Conversely, TLSv1.2 usage decreased from 77.7\% in 2021 represented in light blue in the negative bars to 9.6\% in our dataset represented in navy in the positive bars. The legacy SSL protocols that represented 14.2\% of MAppGraph traffic have been completely eliminated in our captures, indicating industry-wide adoption of secure encryption standards. Apps such as Facebook, Instagram, and Reddit demonstrate this evolution pattern, transitioning from mixed TLS versions to predominantly TLSv1.3 implementations.

The QUIC protocol analysis reveals patterns influenced by both genuine security evolution and the capture system environment. Fig.~\ref{fig:UDP_bihistogram_ourdatasetVSmappgraph} shows the UDP bihistogram with dark blue segments indicating QUIC usage in our dataset in the positive bars and light blue segments representing MAppGraph QUIC traffic in the negative bars. We observe three distinct behaviors of QUIC traffic across both datasets: i) the apps that maintained consistent QUIC behavior, ii) the apps that adopted QUIC traffic and, iii) the apps that generated QUIC traffic in the MAppGraph dataset but show no QUIC traffic in our captures when using mitmproxy.

The apps that maintained consistent QUIC behavior across both datasets such as Chess (com.chess), Facebook Messenger (com.facebook.orca) and Reddit (com.reddit.frontpage). This apps show QUIC traffic in both datasets, represented by both dark blue (our dataset) and light blue (MAppGraph) segments in Fig.~\ref{fig:UDP_bihistogram_ourdatasetVSmappgraph}. The traces of this apps of the MAppGraph dataset show that they already adopted security enhancement using QUIC traffic for their UDP communications back in 2021. 

The apps that have adopted QUIC protocol between 2021 and 2025, representing genuine security enhancements as we expected in the first place. These apps include Instagram (com.instagram.android), SoundCloud (com.soundcloud.android) and Spotify (com.spotify.music). Fig.~\ref{fig:UDP_bihistogram_ourdatasetVSmappgraph} shows these apps with dark blue segments in the positive direction but absent light blue segments in the negative direction. Instagram demonstrates particularly extensive QUIC adoption, showing 19,697 QUIC packets in our dataset compared to zero in MAppGraph. 

Finally, the apps that do not generate QUIC traffic in our dataset appearing without blue coloring in the UDP bihistogram include two subcategories. On the one hand, apps that generated QUIC traffic in the MAppGraph dataset but show no QUIC traffic in our captures when using mitmproxy as Baohay (com.baohay24h.app), TED (com.ted.android) or Noveltoon (mobi.mangatoon.novel.portuguese). 
And, on the other hand, apps that did not generate QUIC traffic in either the MAppGraph dataset or our mitmproxy captures as Among Us (com.innersloth.spacemafia Our), Tivi24h (com.tvonline.tivi24h) or Free Fire (com.dts.freefireth). 
To delve into these apparent regressions and absences, we conducted additional captures without mitmproxy for all these apps and confirmed that they generate QUIC traffic under normal network conditions. This indicates that the absence of QUIC traffic in our dataset results from mitmproxy interception rather than genuine protocol abandonment or lack of QUIC implementation.

Our analysis on the evolution of encrypted protocols demonstrates a clear pivoting toward modern encryption standards in mobile apps. The migration to TLSv1.3, which now represents 90.0\% of TCP encrypted traffic compared to 6.7\% in 2021, indicates widespread adoption of enhanced security protocols across the mobile app ecosystem. This transition eliminates the security vulnerabilities present in legacy SSL protocols and provides improved performance characteristics. And, regarding QUIC protocol adoption, our analysis reveals significant growth in encrypted UDP communications. Under normal network conditions (without mitmproxy), all 50 common apps generate QUIC traffic compared to 29 apps in the MAppGraph dataset, representing a 78.6\% increase in QUIC adoption over the four-year period. However, the capture system environment significantly impacts QUIC traffic visibility, with only 35 apps that generate QUIC traffic when using mitmproxy due to certificate validation mechanisms.

\subsection{DNS Protocol Evolution}\label{dns_evolution}

One of the most significant changes observed between 2021 and 2025 is the evolution of DNS protocols, reflecting the broader industry movement toward privacy-enhanced DNS communications. As presented in Subsection~\ref{Protocol_Overview}, this evolution represents a fundamental shift in how apps resolve domain names.

The MAppGraph dataset from 2021 shows DNS traffic predominantly using unencrypted Do53 protocol, representing 91.0\% of all DNS communications across the 50 common apps, with DoT traffic accounting for only 9.0\%. Although encrypted DNS support was introduced in Android 9~\cite{android_dns_over_tls_blog}, 
where the device automatically upgrades to DoT traffic, this functionality depended on DNS server support. We suppose that the absence of encrypted DNS traffic in the MAppGraph dataset indicates limited server-side support for encrypted DNS protocols in 2021.

In contrast, our 2025 dataset demonstrates a complete transformation of DNS communication patterns. DoT traffic comprises 81.1\% of DNS communications among the common apps, decreasing the usage of Do53. This shift was achieved by configuring our Docker environment to use Google Public DNS servers (8.8.8.8 and 8.8.4.4), which support encrypted versions of DNS traffic. 

\begin{figure}[htbp]
    \centering
    \includegraphics[width=\columnwidth]{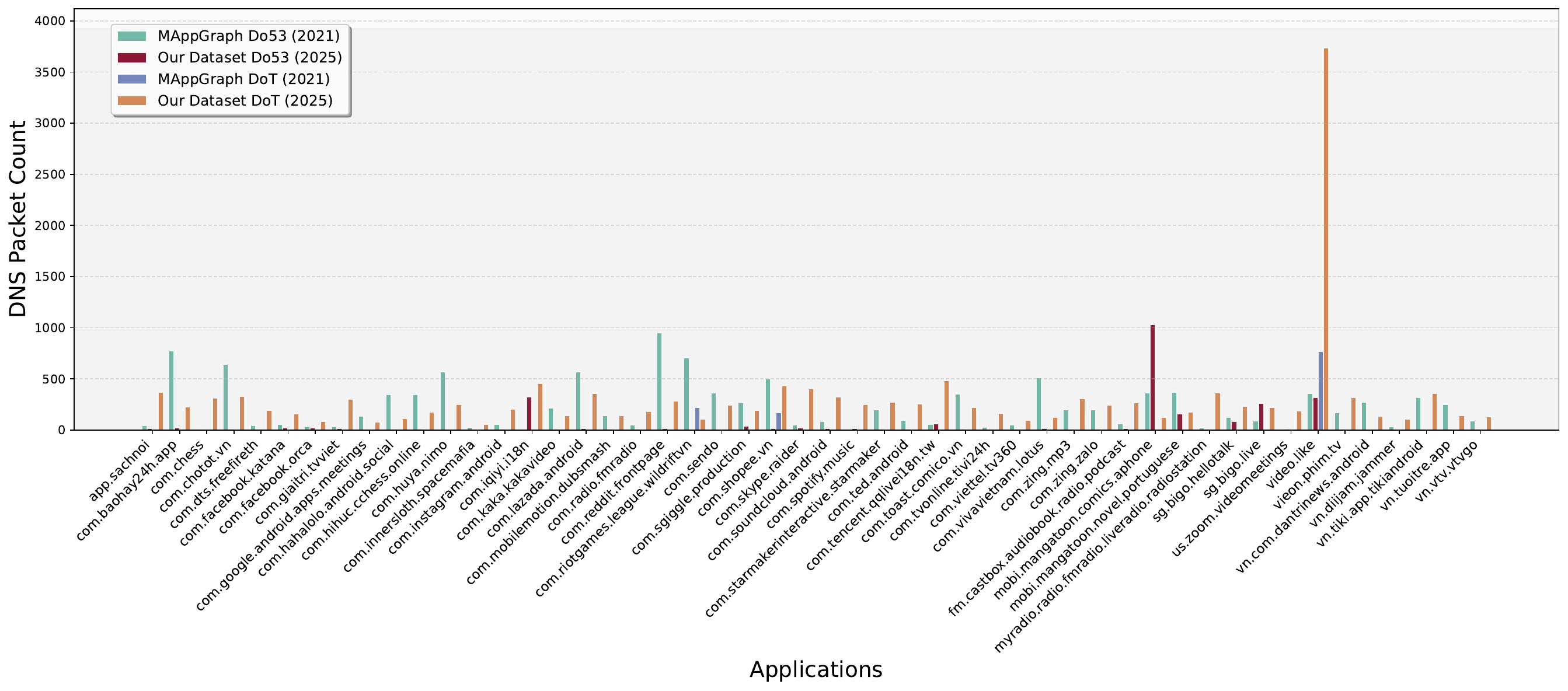}
    \caption{DNS protocol evolution comparison showing Do53 and DoT packet counts across common apps. MAppGraph dataset (2021) shows predominantly Do53 traffic, while our dataset (2025) demonstrates the migration to DoT-encrypted DNS communications.}
    \label{fig:dns_histogram}
\end{figure}

Fig.~\ref{fig:dns_histogram} presents a comparison of Do53 and DoT packet counts across common apps in both datasets. The visualization uses a grouped bar chart format with four distinct categories: MAppGraph Do53 traffic in light blue, our dataset Do53 traffic in burdeaux, MAppGraph DoT traffic in blue, and our dataset DoT traffic in orange. 

We analyze the apps that generated high amounts of residual Do53 traffic. Apps such as \texttt{com.tencent\\.qqlivei18n.tw} and \texttt{mobi.mangatoon.comics.aphone} generate Do53 queries for connectivity checks not only to Google services as observed in Subsection~\ref{Dataset_Characteristics}, but also to CDN domains such as Akamai (\texttt{whatismyip.\\akamai.com}) and other domains, which appear to be connectivity verification mechanisms.

Therefore, the presence of Do53 traffic is related to the mitmproxy capture environment. When capturing traffic without mitmproxy, these apps generate exclusively DoT traffic. The presence of both Do53 and DoT traffic in mitmproxy captures suggests that certain apps implement fallback mechanisms when detecting certificate authority manipulation, reverting to unencrypted DNS queries for specific connectivity checks while maintaining encrypted DNS for primary app functionality.

The evolution from more than 90\% of Do53 traffic in the MAppGraph to more than 80\% of DoT usage in our dataset indicates industry-wide adoption of privacy-enhanced DNS protocols, with apps leveraging encrypted DNS to protect user privacy. 

\section{Conclusion}\label{sec:conclusion}

The rapid evolution of mobile security protocols and the limited availability of recent datasets have constrained researchers' ability to analyze contemporary app behavior and develop robust traffic analysis techniques. This paper addresses two gaps in mobile traffic analysis research: the scarcity of up-to-date app traffic datasets and the absence of tools for systematic traffic capture. 

We present PARROT, a containerized traffic capture system that provides a complete workflow for app traffic collection. The system offers several advantages over existing approaches: reproducibility through containerized deployment of the complete capture system, traffic isolation from host system background traffic, configurable Android version selection to match research requirements, automated labeling and data organization, and optional encryption handling with SSL/TLS key extraction.

Our dataset, collected using the proposed system with mitmproxy enabled, contains traffic from 80 apps with corresponding SSL keys for decryption analysis. The app selection follows the MAppGraph dataset list, with 50 apps in common with the released version of MAppGraph's dataset, enabling direct comparative analysis. Our system provides both encrypted traffic and the corresponding keys for decryption, enabling researchers to analyze both encrypted and decrypted communications.

The comparative analysis between the MAppGraph dataset (2021) and our dataset (2025) reveals significant evolution in app traffic patterns. We observe a migration from TLSv1.2 to TLSv1.3 protocol, with TLSv1.3 comprising 90.0\% of TCP encrypted traffic in 2025 compared to 6.7\% in 2021. QUIC protocol adoption increased substantially, with all 50 common apps generating QUIC traffic under normal network conditions compared to 30 apps in 2021. DNS communications evolved from predominantly unencrypted Do53 protocol to encrypted DNS protocol, representing a fundamental shift toward privacy-enhanced DNS communications. These findings demonstrate the mobile app ecosystem's progression toward enhanced security and privacy implementations.

For future work, we plan to expand our dataset collection to include more apps and app categories, while continuing to improve the capture system's capabilities. The open-source release of PARROT enables the research community to contribute to these efforts and adapt the system for specific research requirements. Additionally, we intent to examine the impact of different capture environments on traffic patterns to provide more insights into app behavior evolution.

\subsection*{Author contributions}
\textbf{Andrea Jimenez-Berenguel:} Data Curation, Investigation, Methodology, Software, Visualization, Writing – original draft.
\textbf{Celeste Campo:} Conceptualization, Supervision, Validation, Writing – review \& editing. 
\textbf{Marta Moure-Garrido:} Supervision, Validation, Writing – review \& editing. 
\textbf{Carlos Garcia-Rubio:} Conceptualization, Validation, Writing – review \& editing. 
\textbf{Daniel Díaz-Sanchez:} Resources, Funding acquisition.
\textbf{Florina Almenares:} Resources, Funding acquisition.

\subsection*{Financial disclosure}
This work has been supported by the Grant DISCOVERY (PID2023-148716OB-C33) funded by MICIU/AEI/10.13039/501100011033 and FEDER, UE; and Grant QURSA (TED2021-130369B-C32) funded by MICIU/AEI/10.13039/501100011033 and European Union NextGenerationEU/PRTR. This work was also supported by Comunidad de Madrid (Spain) under the project RAMONES-CM (TEC2024-COM504), co-financed by European Structural Funds (ESF and FEDER) and  ``I-SHAPER: Internet-Service Hardening of Authentication, Confidentiality, Privacy, Enforcement and Reliability'', from the public invitation program for collaboration in the promotion of Strategic Cybersecurity Projects in Spain - PRTR - of the National Cybersecurity Institute of Spain (INCIBE), this initiative is carried out within the framework of the funds of the Recovery, Transformation and Resilience Plan, financed by the European Union (Next Generation).

\subsection*{Conflict of interest}

The authors declare no potential conflict of interests.

\appendix

\section{App Behavior Under Mitmproxy Interception}
\label{appendix:app_behaviors}

Table~\ref{tab:app_behaviors} presents the behavior of each app when traffic is captured with mitmproxy interception enabled. The table describes the specific limitations and functionalities observed for each application package during the traffic capture process.

{\footnotesize
\captionsetup{margin=2cm} 
\begin{longtable}{p{0.4cm}p{5cm}p{11cm}}
\caption{App behavior under mitmproxy interception during traffic capture}
\label{tab:app_behaviors}\\
\toprule
\textbf{\#} & \textbf{Package Name} & \textbf{App Behavior} \\
\midrule
\endfirsthead
\multicolumn{3}{c}%
{{\bfseries \tablename\ \thetable{} -- continued from previous page}} \\
\toprule
\textbf{\#} & \textbf{Package Name} & \textbf{App Behavior} \\
\midrule
\endhead
\midrule \multicolumn{3}{r}{{Continued on next page}} \\ \midrule
\endfoot
\bottomrule
\endlastfoot
1 & app.sachnoi & launching, navigation through the app but it does not load the content and we cannot login because connection error \\
2 & bbc.mobile.news.ww & launching, navigation through the app but it does not load the content and we cannot login because connection error \\
3 & com.azarlive.android & app freezes at launch \\
4 & com.bachtruyen & launching, navigation through the app but it does not load the content and we cannot login because connection error \\
5 & com.baohay24h.app & launching, navigation through the app but it does not load the content and we cannot login because connection error \\
6 & com.chess & launching, navigation through the app and play off line but cannot play online or login \\
7 & com.chotot.vn & launching, navigation through the app but it does not load the content and we cannot login because connection error \\
8 & com.cnn.mobile.android.phone & launching, navigation through the app but it does not load the content and we cannot login because connection error \\
9 & com.craftbox.jwapp.android & launching and first screen but we cannot browse in the app due to connection errors in the login \\
10 & com.dts.freefireth & app freezes at launch \\
11 & com.facebook.katana & launching and first screen but we cannot browse in the app due to connection errors in the login \\
12 & com.facebook.orca & app freezes at launch \\
13 & com.fplay.activity & launching and the app close automatically because connection error \\
14 & com.giaitri.tvviet & launching, navigation through the app but it does not load the content and we cannot login because connection error \\
15 & com.google.android.apps.meetings & launching and first screen but we cannot browse in the app due to connection errors in the login \\
16 & com.guardian & launching, navigation through the app but it does not load the content and we cannot login because connection error \\
17 & com.hahalolo.android.social & app freezes at launch \\
18 & com.hihuc.cchess.online & launching, navigation through the app but it does not load the content and we cannot login because connection error \\
19 & com.huya.nimo & launching and the app close automatically \\
20 & com.imbb.oversea.android & launching and the app close automatically \\
21 & com.innersloth.spacemafia & launching and the app close automatically \\
22 & com.instagram.android & launching, login and full navigation through the app \\
23 & com.iqiyi.i18n & launching, navigation through the app but it does not load the content and we cannot login because connection error \\
24 & com.kaka.kakavideo & launching, navigation through the app but it does not load the content and we cannot login because connection error \\
25 & com.lazada.android & launching and navigation through the app but it does not load the content because of connection error \\
26 & com.linkedin.android & launching and first screen but we cannot browse in the app due to connection errors in the login \\
27 & com.mobilemotion.dubsmash & launching and first screen but we cannot browse in the app due to connection errors in the login \\
28 & com.netflix.mediaclient & launching and the app close automatically because connection error \\
29 & com.nono.android & launching and the app close automatically because connection error \\
30 & com.pinterest & launching and first screen but we cannot browse in the app due to connection errors in the login \\
31 & com.popsworldwide.popskids & app freezes at launch \\
32 & com.quora.android & app freezes at launch \\
33 & com.radio.fmradio & launching and navigation through the app but it does not load the content because of connection error \\
34 & com.reddit.frontpage & launching and first screen but we cannot browse in the app due to connection errors in the login \\
35 & com.riotgames.league.wildriftvn & app freezes at launch \\
36 & com.sendo & launching, navigation through the app but it does not load the content and we cannot login because connection error \\
37 & com.sgiggle.production & launching and first screen but we cannot browse in the app due to connection errors in the login \\
38 & com.shopee.vn & launching, navigation through the app but it does not load the content and we cannot login because connection error \\
39 & com.skype.raider & launching and first screen but we cannot browse in the app due to connection errors in the login \\
40 & com.snapchat.android & launching and the app close automatically \\
41 & com.soundcloud.android & launching and first screen but we cannot browse in the app due to connection errors in the login \\
42 & com.spotify.music & launching and first screen but we cannot browse in the app due to connection errors in the login \\
43 & com.ss.android.ugc.trill & launching and the app close automatically \\
44 & com.starmakerinteractive.starmaker & launching and first screen but we cannot browse in the app due to connection errors in the login \\
45 & com.ted.android & launching, navigation through the app but it does not load the content and we cannot login because connection error \\
46 & com.tencent.qqlivei18n.tw & launching, navigation through the app but it does not load the content and we cannot login because connection error \\
47 & com.tencent.wesing & app freezes at launch \\
48 & com.tinder & launching and first screen but we cannot browse in the app due to connection errors in the login \\
49 & com.toast.comico.vn & app freezes at launch \\
50 & com.topcv & app freezes at launch \\
51 & com.tvonline.tivi24h & launching, navigation through the app but it does not load the content and we cannot login because connection error \\
52 & com.twitter.android & launching and first screen but we cannot browse in the app due to connection errors in the login \\
53 & com.viettel.tv360 & app freezes at launch \\
54 & com.vivavietnam.lotus & launching and first screen but we cannot browse in the app due to connection errors in the login \\
55 & com.yy.hiyo & launching, first screen and the app close automatically \\
56 & com.zing.mp3 & launching, navigation through the app but it does not load the content and we cannot login because connection error \\
57 & com.zing.zalo & launching and first screen but we cannot browse in the app due to connection errors in the login \\
58 & fm.castbox.audiobook.radio.podcast & launching and first screen but we cannot browse in the app due to connection errors in the login \\
59 & fr.playsoft.vnexpress & launching and navigation through the app but it does not load the content because of connection error \\
60 & io.pobble.sen.android & app freezes at launch \\
61 & me.mycake & launching and the app close automatically because connection error \\
62 & mobi.mangatoon.comics.aphone & launching, navigation through the app but it does not load the content and we cannot login because connection error \\
63 & mobi.mangatoon.novel.portuguese & launching, navigation through the app but it does not load the content and we cannot login because connection error \\
64 & myradio.radio.fmradio.liveradio\newline.radiostation & launching and navigation through the app but it does not load the content because of connection error \\
65 & org.telegram.messenger & launching and first screen but we cannot browse in the app due to connection errors in the login \\
66 & org.wikipedia & launching and navigation through the app but it does not load the content because of connection error \\
67 & sg.bigo.hellotalk & launching and first screen but we cannot browse in the app due to connection errors in the login \\
68 & sg.bigo.live & launching, navigation through the app but cannot login and there are functions of the app that require login \\
69 & tv.twitch.android.app & launching and first screen but we cannot browse in the app due to connection errors in the login \\
70 & us.zoom.videomeetings & launching and first screen but we cannot browse in the app due to connection errors in the login \\
71 & vcc.mobilenewsreader.sohanews & app freezes at launch \\
72 & video.like & launching, navigation through the app but cannot login and there are functions of the app that require login \\
73 & vieon.phim.tv & app freezes at launch \\
74 & vn.com.dantrinews.android & launching, navigation through the app but it does not load the content and we cannot login because connection error \\
75 & vn.diijam.jammer & launching and first screen but we cannot browse in the app due to connection errors in the login \\
76 & vn.tiki.app.tikiandroid & launching, navigation through the app but it does not load the content and we cannot login because connection error \\
77 & vn.tuoitre.app & launching, navigation through the app but it does not load the content and we cannot login because connection error \\
78 & vn.vtv.vtvgo & app freezes at launch \\
79 & wsj.reader\_sp & launching, navigation through the app but it does not load the content and we cannot login because connection error \\
80 & xyz.wmfall.phim & app freezes at launch \\
\end{longtable}
}

\bibliographystyle{unsrt} 
\bibliography{references} 

\begin{thebibliography}{10}

\bibitem{rfc1035dns}
{Domain names - implementation and specification}.
\newblock RFC 1035, November 1987.

\bibitem{rfc7858dot}
Zi~Hu, Liang Zhu, John Heidemann, Allison Mankin, Duane Wessels, and Paul~E. Hoffman.
\newblock {Specification for DNS over Transport Layer Security (TLS)}.
\newblock RFC 7858, May 2016.

\bibitem{rfc8484doh}
Paul~E. Hoffman and Patrick McManus.
\newblock {DNS Queries over HTTPS (DoH)}.
\newblock RFC 8484, October 2018.

\bibitem{rfc9250doq}
Christian Huitema, Sara Dickinson, and Allison Mankin.
\newblock {DNS over Dedicated QUIC Connections}.
\newblock RFC 9250, May 2022.

\bibitem{rfc9230odoh}
Eric Kinnear, Patrick McManus, Tommy Pauly, Tanya Verma, and Christopher~A. Wood.
\newblock {Oblivious DNS over HTTPS}.
\newblock RFC 9230, June 2022.

\bibitem{android_cleartext_communications}
{Android Developers}.
\newblock Cleartext communications.
\newblock \url{https://developer.android.com/privacy-and-security/risks/cleartext-communications}, 2024.
\newblock Accessed: 2025.

\bibitem{android_security_ssl_tls13}
{Android Developers}.
\newblock Security with network protocols.
\newblock \url{https://developer.android.com/privacy-and-security/security-ssl\#tls-1.3-enabled-by-default}, 2025.
\newblock Section: TLS 1.3 enabled by default. Accessed: 2025.

\bibitem{android_network_stacks}
{Android Developers}.
\newblock Network stacks.
\newblock \url{https://developer.android.com/media/media3/exoplayer/network-stacks\#supported-network}, 2024.
\newblock Section: Supported network stacks. Accessed: 2025.

\bibitem{android_insecure_dns}
{Android Developers}.
\newblock Insecure dns setup.
\newblock \url{https://developer.android.com/privacy-and-security/risks/bad-dns\#risk:-vulnerable-dns-transport-security}, 2024.
\newblock Section: Risk: Vulnerable DNS Transport Security. Accessed: 2025.

\bibitem{Mankowski2023dataset}
Dimitri Mankowski, Thom Wiggers, and Veelasha Moonsamy.
\newblock Tls → post-quantum tls: Inspecting the tls landscape for pqc adoption on android.
\newblock In {\em 2023 IEEE European Symposium on Security and Privacy Workshops (EuroS\&PW)}, pages 526--538, July 2023.

\bibitem{Pham2021mappgraph}
Thai-Dien Pham, Thien-Lac Ho, Tram Truong-Huu, Tien-Dung Cao, and Hong-Linh Truong.
\newblock {MAppGraph}: Mobile-app classification on encrypted network traffic using deep graph convolution neural networks.
\newblock In {\em Proceedings of the 37th Annual Computer Security Applications Conference}, ACSAC '21, page 1025–1038, New York, NY, USA, 2021. Tan Tao University, Singapore Institute of Technology, Aalto University.

\bibitem{Zhao2022NUDTdataset}
Shuang Zhao, Jincheng Zhong, Shuhui Chen, and Jianbing Liang.
\newblock Comprehensive mobile traffic characterization based on a large-scale mobile traffic dataset.
\newblock In Xingliang Yuan, Guangdong Bai, Cristina Alcaraz, and Suryadipta Majumdar, editors, {\em Network and System Security}, pages 214--232, Cham, 2022. National University of Defense Technology, Changsha, 410073, China, Springer Nature Switzerland.

\bibitem{Bayat2024ITC60dataset}
Marziyeh Bayat, Javad Garshasbi, Mozhgan Mehdizadeh, Neda Nozari, Abolghasem Rezaei~Khesal, Maryam Dokhaei, and Mehdi Teimouri.
\newblock Itc-net-blend-60: a comprehensive dataset for robust network traffic classification in diverse environments.
\newblock {\em BMC Research Notes}, 17(1):165, 2024.

\bibitem{Wang2020netlog}
Xin Wang, Shuhui Chen, and Jinshu Su.
\newblock Real network traffic collection and deep learning for mobile app identification.
\newblock {\em Wireless Communications and Mobile Computing}, 2020(1):4707909, 2020.

\bibitem{Rezaei2020notpublicdataset}
Shahbaz Rezaei, Bryce Kroencke, and Xin Liu.
\newblock Large-scale mobile app identification using deep learning.
\newblock {\em IEEE Access}, 8:348--362, 2020.

\bibitem{solost}
Johnny So, Iskander Sanchez-Rola, and Nick Nikiforakis.
\newblock Lost in the mists of time: Expirations in dns footprints of mobile apps.
\newblock Accessed: 2025.

\bibitem{Ren2019dataset}
Jingjing Ren, Daniel~J. Dubois, and David Choffnes.
\newblock An international view of privacy risks for mobile apps.
\newblock https://recon.meddle.mobi/papers/cross-market.pdf, 2019.

\bibitem{Aceto2019miragedataset}
Giuseppe Aceto, Domenico Ciuonzo, Antonio Montieri, Valerio Persico, and Antonio Pescapé.
\newblock Mirage: Mobile-app traffic capture and ground-truth creation.
\newblock In {\em 2019 4th International Conference on Computing, Communications and Security (ICCCS)}, pages 1--8, Oct 2019.

\bibitem{jimenez2025parrot}
Andrea Jimenez-Berenguel, Celeste Campo, Marta Moure-Garrido, and Carlos Garcia-Rubio.
\newblock {PARROT2025\_mitmproxy Dataset: Mobile App Traffic Traces with SSL Keys}.
\newblock July 2025.
\newblock Dataset.

\bibitem{Nikbakht2024ITC5dataset}
Mohammad Nikbakht and Mehdi Teimouri.
\newblock Itc-net-audio-5: an audio streaming dataset for application identification in network traffic classification.
\newblock {\em BMC Research Notes}, 17(1):57, 2024.

\bibitem{Jiang2025dataset}
Wei Jiang, Bin Zhang, Qixun Zhu, Conghui Liao, and Wenyong Wang.
\newblock A real network environment dataset for traffic analysis.
\newblock {\em Scientific Data}, 12(1):756, 2025.

\bibitem{Jiang2023dataset}
Minghao Jiang, Mingxin Cui, Chang Liu, Gaopeng Gou, Gang Xiong, and Zhen Li.
\newblock Zero-relabelling mobile-app identification over drifted encrypted network traffic.
\newblock {\em Computer Networks}, 228:109728, 2023.

\bibitem{Guarino2024datasetmirage}
Idio Guarino, Domenico Ciuonzo, Antonio Montieri, and Antonio Pescapè.
\newblock Mirage-app×act-2024: A novel dataset for mobile app and activity traffic analysis.
\newblock In {\em 2024 20th International Conference on Wireless and Mobile Computing, Networking and Communications (WiMob)}, pages 663--666, Oct 2024.

\bibitem{Taylor2016appscanner}
Vincent~F. Taylor, Riccardo Spolaor, Mauro Conti, and Ivan Martinovic.
\newblock Appscanner: Automatic fingerprinting of smartphone apps from encrypted network traffic.
\newblock In {\em 2016 IEEE European Symposium on Security and Privacy (EuroS\&P)}, pages 439--454, March 2016.

\bibitem{faranda2025pcapdroid}
Emanuele Faranda.
\newblock Pcapdroid: No-root network monitor, firewall and pcap dumper for android.
\newblock GitHub repository, 2025.

\bibitem{mitmproxy}
Aldo Cortesi, Maximilian Hils, and Thomas Kriechbaumer.
\newblock mitmproxy.
\newblock \url{https://mitmproxy.org/}.

\bibitem{androidsdk}
{Android Developers}.
\newblock Android sdk command-line tools.
\newblock \url{https://developer.android.com/studio}.

\bibitem{android_captive_portal}
{Android Developers}.
\newblock Captive portal api support.
\newblock \url{https://developer.android.com/about/versions/11/features/captive-portal}, 2025.
\newblock Accessed: 2025.

\bibitem{android_dns_over_tls_blog}
Erik Kline and Ben Schwartz.
\newblock Dns over tls support in android p developer preview.
\newblock Android Developers Blog. \url{https://android-developers.googleblog.com/2018/04/dns-over-tls-support-in-android-p.html}, 2018.
\newblock Accessed: 2025.

\end{thebibliography}

\subsection*{Author Biography}

\vspace{1em}
\noindent

\begin{wrapfigure}{l}{1.2in}
\vspace{-20pt}
\includegraphics[width=1.1in,height=1.3in,keepaspectratio]{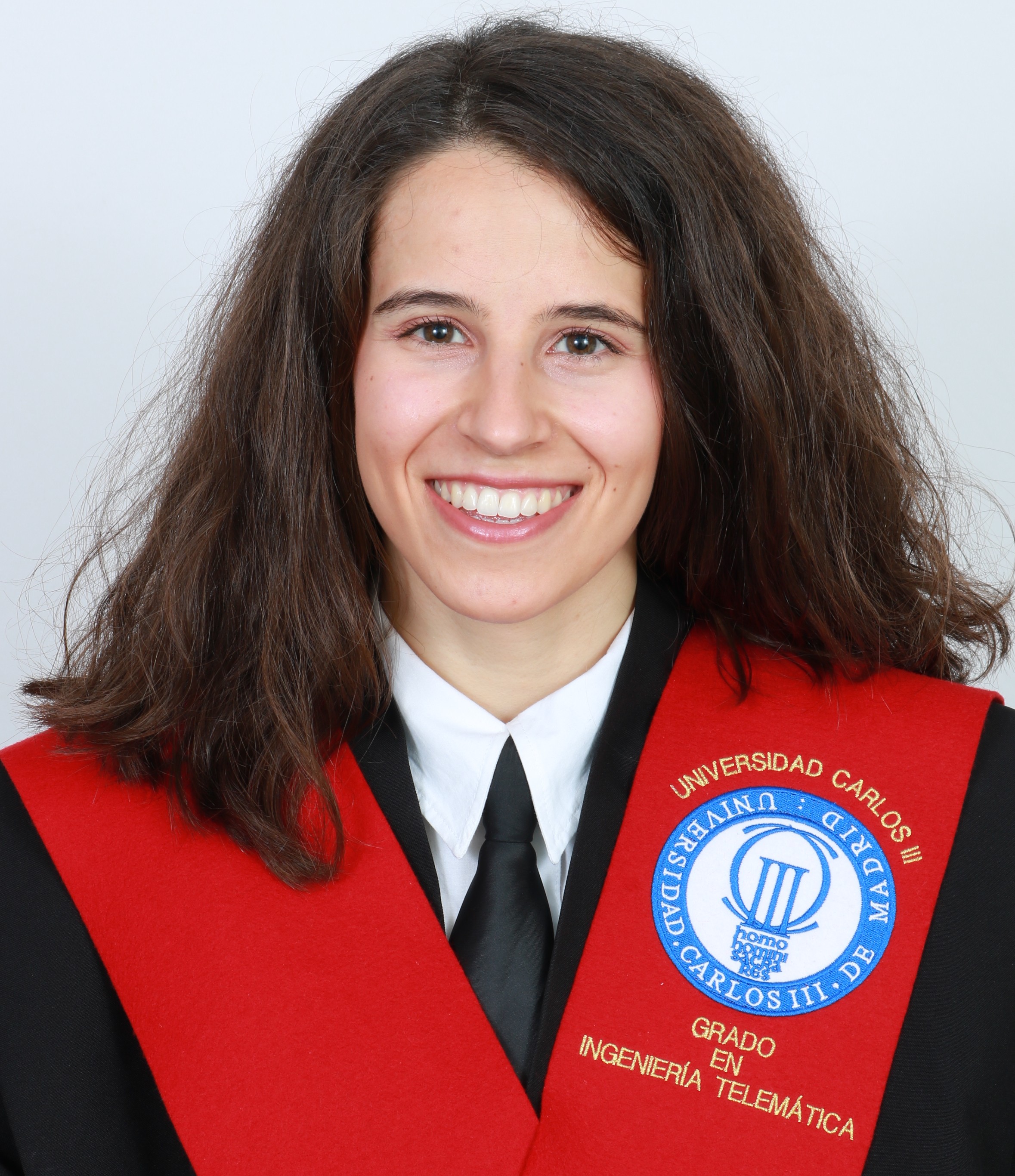}
\vspace{-30pt}
\end{wrapfigure}
\noindent \textbf{Andrea Jimenez-Berenguel} is a PhD Student at the Department of Telematic Engineering of the Universidad Carlos III de Madrid. Her research journey is focused on Android traffic analysis and user privacy. She received her bachelor degree in Telematic Engineering in 2022 and her MS degree in Telecommunications Engineering in 2024, both from the Universidad Carlos III de Madrid.

\vspace{4em}

\begin{wrapfigure}{l}{1.2in}
\vspace{-20pt}
\includegraphics[width=1.1in,height=1.3in,keepaspectratio]{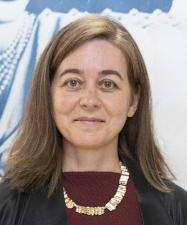}
\vspace{-40pt}
\end{wrapfigure}
\noindent \textbf{Celeste Campo} received her Ph.D. degree from the University Carlos III of Madrid in 2004. She is an associate professor at the Department of Telematic Engineering of the University Carlos III of Madrid. Her research interests include design and performance evaluation of communication protocols for ad hoc networks, energy-aware communications, and middleware technologies for pervasive computing.

\vspace{4em}

\begin{wrapfigure}{l}{1.2in}
\vspace{-20pt}
\includegraphics[width=1.1in,height=1.3in,keepaspectratio]{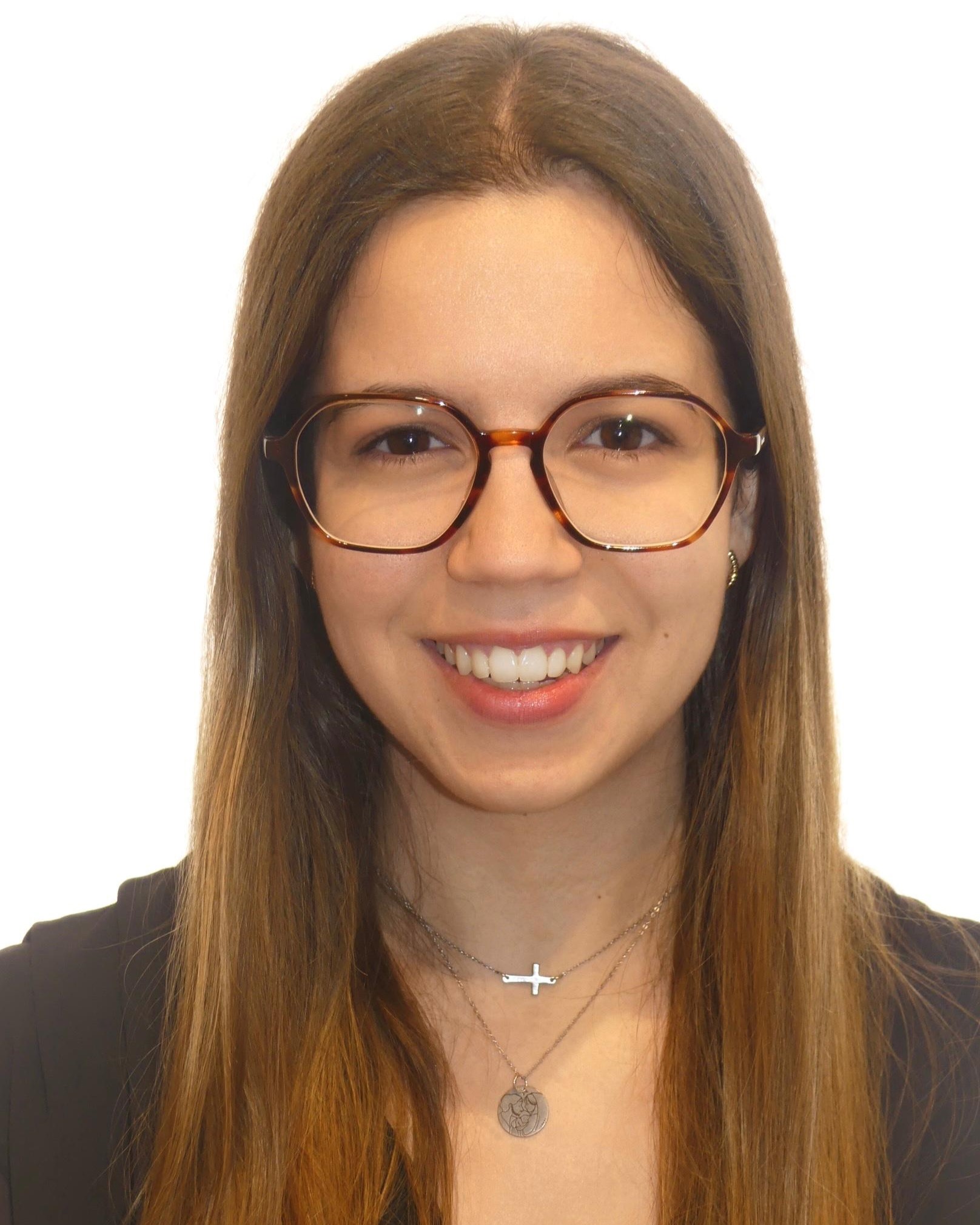}
\vspace{-50pt}
\end{wrapfigure}
\noindent \textbf{Marta Moure-Garrido} is an Assistant Professor in the Department of Telematic Engineering of the University Carlos III of Madrid. Her research focuses on communication protocol design, network security, and privacy-preserving network monitoring. She received her Ph.D. degree from the University Carlos III of Madrid in 2024.

\vspace{5em}

\begin{wrapfigure}{l}{1.2in}
\vspace{-20pt}
\includegraphics[width=1.5in,height=1.3in,keepaspectratio]{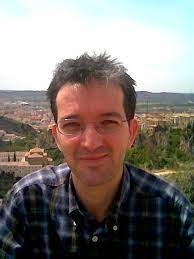}
\vspace{-40pt}
\end{wrapfigure}
\noindent \textbf{Carlos Garcia-Rubio} received the Ph.D. degree from the Technical University of Madrid in 2000. He is an associate professor at the Department of Telematic Engineering of the University Carlos III of Madrid. His research focus is centered on mobile and wireless networked computing systems, and on the design and performance evaluation of communication protocols, mainly at the transport and application layers.

\vspace{5em}

\begin{wrapfigure}{l}{1.2in}
\vspace{-20pt}
\includegraphics[width=1.1in,height=1.3in,keepaspectratio]{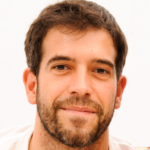}
\vspace{-10pt}
\end{wrapfigure}
\noindent \textbf{Daniel Díaz-Sanchez} holds a Ph.D. in Telematics Engineering from University Carlos III of Madrid (UC3M), where he's now an Associate Professor. His research focuses on distributed authentication/authorization, data/content protection, distributed computing, and the Internet of Things (IoT). He's received several distinctions, including UC3M's Extraordinary Ph.D. Award (2009) and the Best Ph.D. Award from the Spanish Association of Telecommunications Engineers (2009). 
His contributions have also been recognized by the IEEE with the Chester Sall Award (2012), Senior Member elevation (2014), and an Appreciation Certificate for outstanding editorial efforts in Transactions on Consumer Electronics (2022). Beyond his research, Daniel is deeply involved in internationalization and publishing. He actively organized Consumer Technology conferences, has served as an Associate Editor at IEEE for over a decade, founded and managed the IEEE Spanish Chapter of Consumer Technology as its VP, and supervises numerous research projects and academic works in his field.

\vspace{1em}

\begin{wrapfigure}{l}{1.2in}
\vspace{-20pt}
\includegraphics[width=1.2in,height=1.5in,keepaspectratio]{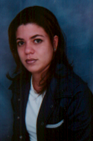}
\vspace{-10pt}
\end{wrapfigure}
\noindent \textbf{Florina Almenares} received the M.Sc. degree in Telematic and the Ph.D. degree from the University Carlos III of Madrid, in 2003 and 2006, respectively. Since 2008, she has worked as an Associate Professor with the Department of Telematic Engineering, University Carlos III of Madrid. Her research interests include post-quantum cryptography (PQC), cybersecurity, machine learning, trust and reputation management models, identity management, secure architectures, and risk assessment. This research has been recently applied to ubiquitous computing and IoT, smart grids, and smart cities. She received the IEEE Chester Sall Award (2012). She has experience on PQC and has supervised research projects and academic works in this field.

\end{document}